\documentclass[10pt]{elsarticle}
\usepackage[utf8]{inputenc}
\usepackage[margin=3cm]{geometry}
\usepackage{amsmath}
\usepackage{graphicx}
\usepackage{framed} 
\usepackage[labelfont=bf]{caption}
\usepackage[title]{appendix}
\usepackage{subfigure}
\usepackage{wrapfig}
\usepackage{multicol}
\usepackage{subfloat}
\usepackage{floatrow}
\usepackage{caption}
\DeclareCaptionType{equ}[][]
\usepackage{xcolor}
\usepackage{siunitx}
\usepackage{gensymb}
\usepackage[parfill]{parskip}
\usepackage{textcomp}
\usepackage{mathtools}
\usepackage{color}
\usepackage{longtable}
\usepackage{listings}
\usepackage{enumitem}

\usepackage{indentfirst}
\usepackage{booktabs}
\usepackage{tabularx}
\definecolor{codegreen}{rgb}{0,0.6,0}
\definecolor{codegray}{rgb}{0.5,0.5,0.5}
\definecolor{codepurple}{rgb}{0.58,0,0.82}
\definecolor{backcolour}{rgb}{0.95,0.95,0.92}
\lstdefinestyle{mystyle}{
    backgroundcolor=\color{backcolour},   
    commentstyle=\color{codegreen},
    keywordstyle=\color{magenta},
    numberstyle=\tiny\color{codegray},
    stringstyle=\color{codepurple},
    basicstyle=\ttfamily\footnotesize,
    breakatwhitespace=false,         
    breaklines=true,                 
    captionpos=b,                    
    keepspaces=true,                                     
    numbersep=5pt,                  
    showspaces=false,                
    showstringspaces=false,
    showtabs=false,                  
    tabsize=4
}
\newcolumntype{P}[1]{>{\centering\arraybackslash}p{#1}}
\graphicspath{{fig/}}
\usepackage[breaklinks=true,colorlinks=true,linkcolor=blue,urlcolor=blue,citecolor=blue]{hyperref}
\biboptions{sort&compress}
\usepackage{cleveref}
\usepackage{physics}
\newcommand{\blue}[1]{{\color{blue}#1}} 

\makeatletter
\def\ps@pprintTitle{%
  \let\@oddhead\@empty
  \let\@evenhead\@empty
  \def\@oddfoot{\reset@font\hfil\thepage\hfil}
  \let\@evenfoot\@oddfoot
}
\newcommand{\myfootnote}[1]{
    \renewcommand{\thefootnote}{}
    \footnotetext{\scriptsize#1}
    \renewcommand{\thefootnote}{\arabic{footnote}}
}
\makeatother
\usepackage{setspace}

\begin{document}

\begin{frontmatter}

\title{Feedforward equilibrium trajectory optimization with GSPulse}
\author[CFS]{J.T. Wai$^{*,}$}
\author[CFS]{M.D. Boyer$^{**,}$}
\author[CFS]{D.J. Battaglia}
\author[EPFL]{A. Merle}
\author[EPFL]{F. Carpanese}
\author[GDM]{F. Felici}
\author[CCFE]{M. Kochan}
\author[Princeton,PPPL]{E. Kolemen} 

\address[CFS]{Commonwealth Fusion Systems, Devens, MA, USA}
\address[EPFL]{École Polytechnique Fédérale de Lausanne, Lausanne, Switzerland}
\address[GDM]{Google Deepmind, London, UK}
\address[CCFE]{Culham Centre for Fusion Energy, Abingdon, United Kingdom}
\address[Princeton]{Princeton University, Princeton, New Jersey, USA}
\address[PPPL]{Princeton Plasma Physics Laboratory, Princeton, New Jersey, USA}

\begin{abstract}
\indent 

One of the common tasks required for designing new plasma scenarios or evaluating capabilities of a tokamak is to design the desired equilibria using a Grad-Shafranov (GS) equilibrium solver. However, most standard equilibrium solvers are time-independent and do not include dynamic effects such as plasma current flux consumption, induced vessel currents, or voltage constraints. Another class of tools, plasma equilibrium evolution simulators, do include time-dependent effects. These are generally structured to solve the forward problem of evolving the plasma equilibrium given feedback-controlled voltages. In this work, we introduce GSPulse, a novel algorithm for equilibrium trajectory optimization, that is more akin to a pulse planner than a pulse simulator. GSPulse includes time-dependent effects and solves the inverse problem: given a user-specified set of target equilibrium shapes, as well as limits on the coil currents and voltages, the optimizer returns trajectories of the voltages, currents, and achievable equilibria. This task is useful for scoping performance of a tokamak and exploring the space of achievable pulses. The computed equilibria satisfy both Grad-Shafranov force balance and axisymmetric circuit dynamics. The optimization is performed by restructuring the free-boundary equilibrium evolution (FBEE) equations into a form where it is computationally efficient to optimize the entire dynamic sequence. GSPulse can solve for hundreds of equilibria simultaneously within a few minutes. GSPulse has been validated against NSTX-U and MAST-U experiments and against SPARC feedback control simulations, and is being used to perform scenario design for SPARC. The computed trajectories can be used as feedforward inputs that are connected to the feedback controller to inform and improve feedback performance. The code for GSPulse is available open-source at \blue{github.com/jwai-cfs/GSPulse\_public}.

\end{abstract}
\end{frontmatter}

\myfootnote{Corresponding authors: jwai@cfs.energy, ekolemen@princeton.edu}
\myfootnote{$^*$ Portions of this work were completed while affiliated with Princeton University}
\myfootnote{$^{**}$ Portions of this work were completed while affiliated with Princeton Plasma Physics Laboratory}


\section{Introduction}

\crefname{appendix}{}{}

\subsection{Motivation}

During a tokamak pulse, time-varying voltages and currents are applied to coils within the tokamak, which drive plasma current and create magnetic fields that confine the plasma. The shape and distribution of these magnetic field surfaces is referred to as the plasma equilibrium, and one of the common tasks required for designing new plasma pulses or scenarios is to compute these equilibria using a Grad-Shafranov equilibrium solver. When evaluating a particular scenario it can be important to consider characteristics of the equilibrium such as: the overall shape that can be achieved, the location of the strike point legs which influences power exhaust, what the required coil currents are, and what range of plasma current or other core plasma properties are feasible given the device hardware constraints. This type of equilibrium analysis is critical for defining boundaries of the machine's operational envelope. 

With feedforward trajectory design, we aim to solve for time-dependent sequences of plasma equilibria since there are a variety of time-dependent effects that are not captured by static equilibrium analysis and can only be captured with dynamic models. One example is the flux transformer effect, where the plasma current must be sustained by continuously evolving the other coils to create an electric field. As the coil currents are ramped throughout the pulse to provide this electric field they will eventually reach their hardware limits. This influences the range of equilibria that are achievable at different times during the pulse. Another dynamic effect is the evolution of currents that are induced in the tokamak's vacuum vessel conducting structures. The magnitude and distribution of these vessel currents are both caused by the equilibrium evolution and play a role in modifying the evolution. Lastly, the impact of power supply voltage limits on the equilibrium is a dynamic effect. The voltages constrain the speed at which the currents and equilibrium can evolve throughout the pulse. 

The GSPulse code implements a novel algorithm for designing feedforward equilibrium trajectories which is useful for a variety of operational and design tasks. For example: 

\begin{itemize}[itemsep=0.5em]

    \item Evaluating the operational envelope of a machine, such as what pulses are achievable given constraints on the coil voltages, currents, and ohmic flux supply.  
    
     \item Designing new scenarios and reference targets for a pulse. Since not every requested plasma shape or scenario is achievable, it is important to design reasonable reference targets so that the feedback controller is not requested to drive the system along unsuitable trajectories. 
     
    \item On many tokamaks, portions of the pulse routinely use only feedforward actuation when feedback control is difficult or unavailable. For example, the sequence for plasma initiation usually includes driving the coils along preprogrammed trajectories to apply large voltages for creating plasma breakdown conditions and sustaining plasma burnthrough. This is often done in feedforward-only before transitioning to feedback control. Some reasons for this are that during the highly dynamic early portion of the pulse the phase delay associated with feedback control might be unacceptably high, and that, at low plasma current the signal-to-noise ratio is lower and some real-time reconstruction algorithms may not converge reliably \cite{Wai2022} (although simpler non-Grad-Shafranov algorithms could suffice). 

    \item Improving the feedback control performance. There are several ways that feedback control can be improved with complementary application of feedforward. For example, feedback phase delays can be reduced by applying voltage feedforward, improving robustness or other control properties. Another example is that the feedback controller may have extra degrees of freedom, since often multiple different coil current combinations can lead to approximately the same shape. We can use current feedforward to bias the feedback control along advantageous trajectories, for example, trajectories that: are smooth in time, require less actuator effort, avoid imbalanced currents or coils ``fighting'' each other, or have more margin to reaching hardware constraints (e.g. force limits). 

\end{itemize}

\begin{figure}[H] 
    \begin{center}
        \includegraphics[width=\linewidth]{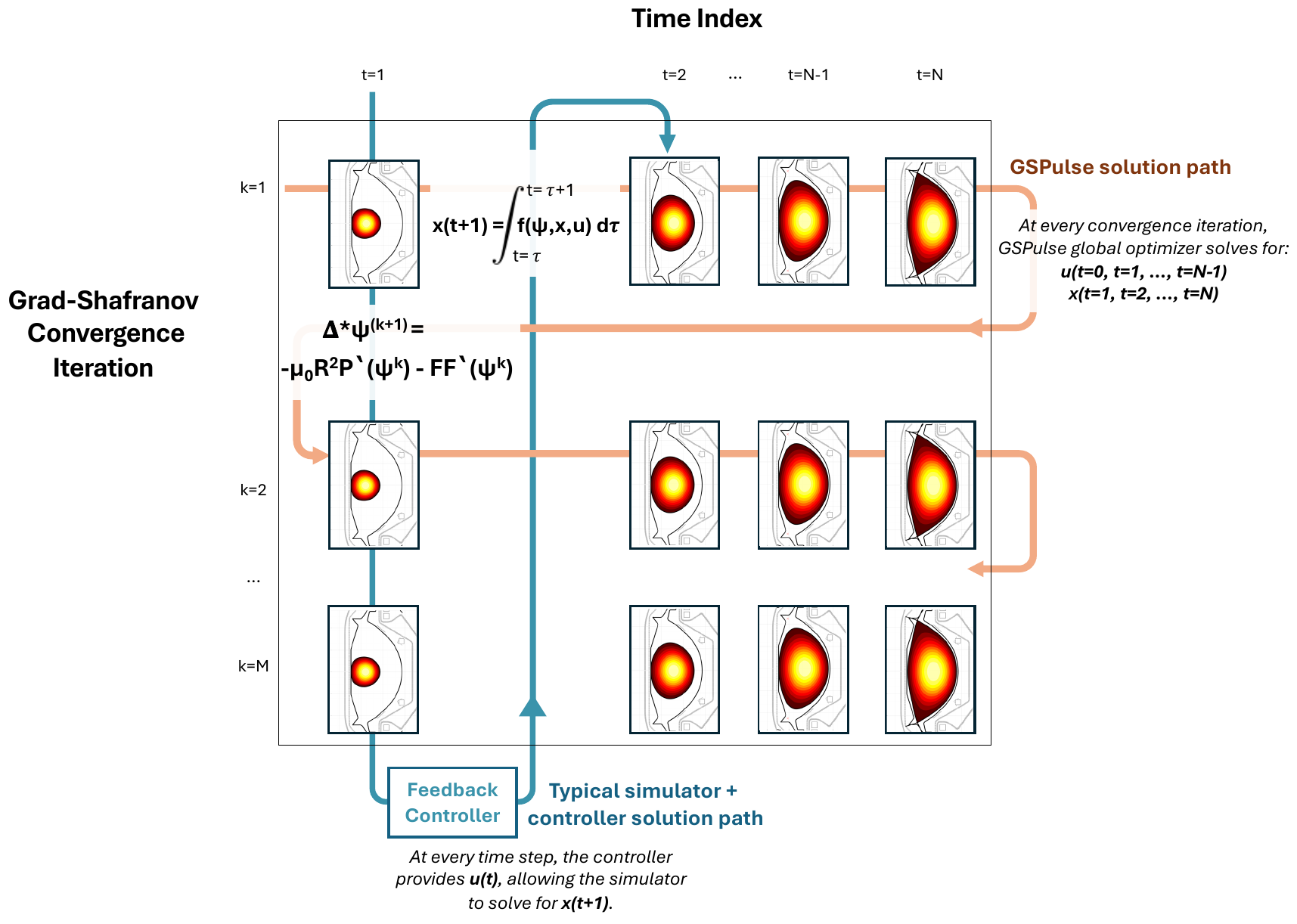}
    \end{center}
    \caption{Schematic of the GSPulse algorithm. A sequence of equilibria satisfy the free-boundary equilibrium evolution (FBEE) model if each equilibrium is a fully converged Grad-Shafranov solution and if equilibria are linked in time by the appropriate dynamics. These two conditions are represented by the two axes dimensions, time and Grad-Shafranov convergence. A typical approach for pulse simulating is to converge an equilibrium, determine feedback voltage commands, and evolve the system one step forward in time and solve for the next equilibrium. GSPulse re-structures the solution order. Beginning with a rough, unconverged estimate of an equilibrium sequence, GSPulse solves a global optimization problem for the full time evolution of the conductor currents $x(t)$ and voltages $u(t)$. Then, the algorithm takes a step along the convergence dimension, performing a Picard iteration of the equilibrium flux for all time slices. By rearranging the problem structure in this form, the GSPulse optimizer can perform global path-planning for time-dependent tasks such as avoiding hardware constraints or identifying smooth trajectories.}
    \label{fig:gspulse_convergence}
\end{figure}

\subsection{Pulse simulating vs pulse planning}

One important point to note is the distinction between pulse simulating and pulse planning (see \cref{tab:equilibrium_solvers}). While both classes of algorithms follow the same governing equations for free-boundary equilibrium evolution (FBEE), they do so with different methodologies and application purposes. A pulse simulator, sometimes called the forward evolutive problem, evolves the equilibrium step-by-step given a set of voltages received from a feedback controller in the loop. By contrast, the GSPulse pulse planner solves the inverse evolutive problem of computing trajectories of voltages, currents, and equilibria given a set of target equilibrium features. It finds an equilibrium evolution that best matches a set of user-specified targets. This is fundamentally a global problem in that we need to optimize the trajectory considering the entirety of the evolution, which modifies the Grad-Shafranov convergence strategy (\cref{fig:gspulse_convergence}) discussed in the next section. 

\begin{table}[H] 
    \begin{center}
        \includegraphics[width=0.7\linewidth]{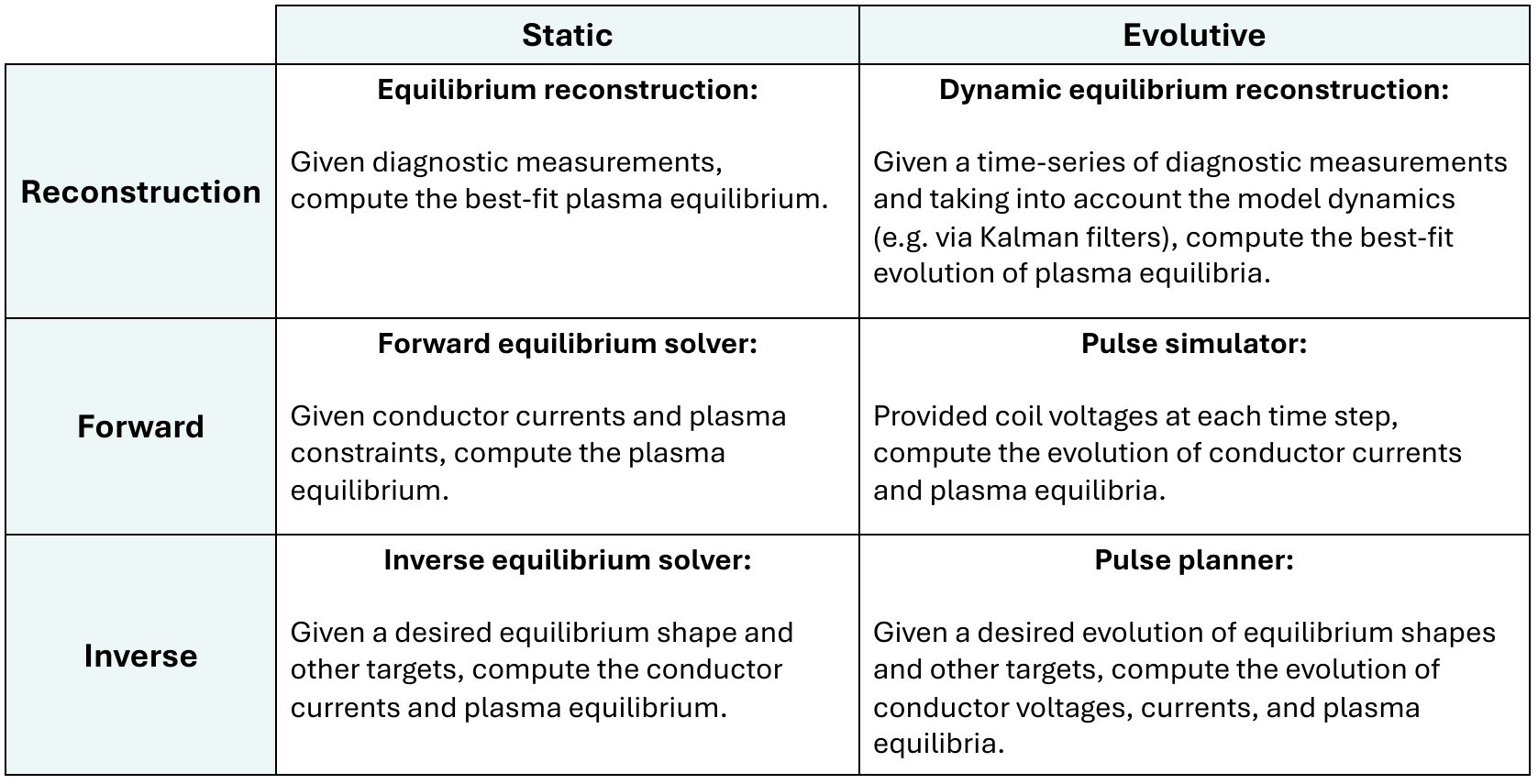}
    \end{center}        
    \caption{A classification scheme for various types of equilibrium algorithms, although it is noted that not all equilibrium codes necessarily fit cleanly into this schema and might combine features of multiple categories. GSPulse is an ``evolutive inverse'' equilibrium solver for use in pulse planning, and when solving for a single equilibrium, collapses into a static inverse solver.}
    \label{tab:equilibrium_solvers}
\end{table}

An additional difference is that, due to the convergence strategy and inherent vertical stabilization (discussed further in \cref{sec:vs_convergence}), the GSPulse pulse planner can take larger time steps than a pulse simulator while remaining vertically stable. This allows for a reduced computational cost of the solution. Some notable examples of forward evolutive pulse simulators are DINA \cite{Khayrutdinov1993}, FGE \cite{carpanese2021}, FreeGSNKE \cite{Amorisco2024}, and GSevolve \cite{Welander2019}.

\subsection{Comparison against previous approaches}

The core idea of the GSPulse algorithm is that it rearranges the structure of the FBEE model to enable performing a time-dependent optimization over the entire sequence of equilibria. This improves on some other methods for designing feedforward trajectories. While there is no standard method for defining these trajectories, one of the typical approaches would be to first design a series of static equilibria and then go back and modify these equilibria with an attempt at compensating for the dynamic effects such as induced vessel currents. One of the downsides to this approach is that it can be computationally expensive, since each of the equilibria must fully converge more than once. Additionally, while the equilibria may have the appropriate dynamic linking, it is still difficult to tune the trajectory into an optimal trajectory, since all of the optimization comes from user-effort in defining the sequence of equilibria, and not from an explicit trajectory optimization. 


Several earlier solvers have also treated the inverse-evolutive optimization problem, including CEDRES++ \cite{Heumann2015} and an algorithm by Blum \cite{Blum2019}, which have now been incorporated into the NICE code \cite{Faugeras2020}. These solvers use Sequential Quadratic Programming (SQP) methods to optimize for sequences of equilibria performing a similar task as GSPulse. The main differences of these solvers versus GSPulse are: the exact formulation of the cost functional where GSPulse includes additional terms on the derivatives and smoothness of trajectories, and that the SQP-based solvers iterate on the full state vector solution directly, whereas GSPulse splits the iteration into optimizing the vacuum conductor evolution and Picard iterations of the plasma state. This framework allows GSPulse to re-use sensitivity matrices used in the optimization, and forgo additional gradient or adjoint state calculations. The implicit assumption is that it is sufficient to simply update the effect of the plasma contribution to the cost function at each iteration, and not necessary to use the gradient of the cost w.r.t plasma contribution within the optimization step. This structure is very beneficial for computational efficiency although perhaps reduces the optimality of the final solution. 

Another code within this design space is Tokamaker \cite{Hansen2024}, which also includes some time-dependent effects and can solve for sequences of equilibria, but without performing global trajectory optimization.  Additionally, the FBT \cite{Hofmann1988} equilibrium code has recently been extended to treat the dynamic case.

\section{Algorithm principles}\label{sec:gspulse}

\begin{figure}[H] 
    \begin{center}
        \makebox[\textwidth][c]{\includegraphics[width=0.9\linewidth]{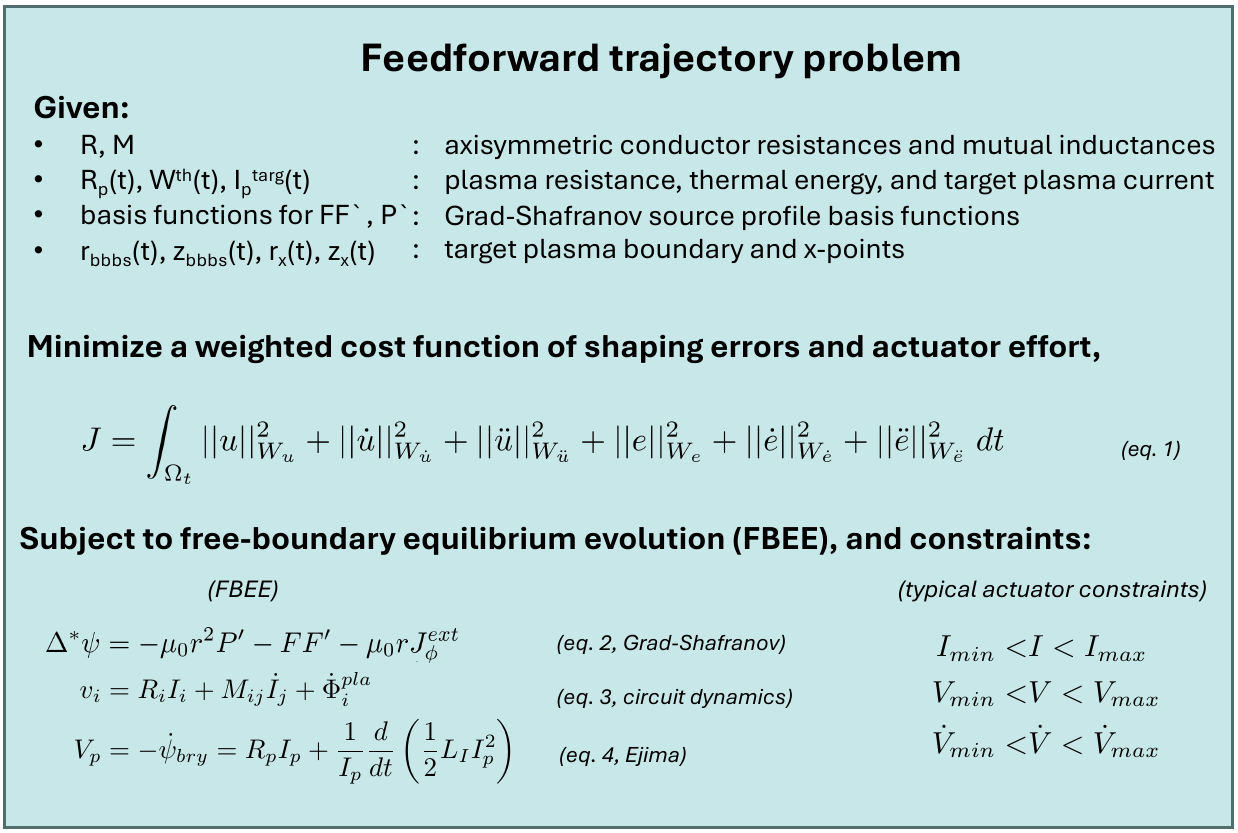}}
    \end{center}
    \caption{Summary of the feedforward trajectory optimization problem. GSPulse minimizes a quadratic cost function of the isoflux shaping errors and actuator effort, subject to the free-boundary equilibrium evolution (FBEE) governing equations, and also subject to any power supply constraints.}
    \label{fig:ff}
\end{figure}

This section describes the working principles for the GSPulse algorithm. The full derivation is provided in \cref{app:gspulse_algorithm}. The premise is to minimize a cost function associated with the shaping evolution, while simultaneously satisfying the FBEE governing equations and actuator constraints. The cost function used is: 

\begin{equation}
    J = \int_{\Omega_t} ||u||^2_{W_u} +  ||\dot u||^2_{W_{\dot u}} + ||\ddot u||^2_{W_{\ddot u}} + ||e||^2_{W_e} +  ||\dot e||^2_{W_{\dot e}} + ||\ddot e||^2_{W_{\ddot e}} \; dt
\end{equation}

In this equation, $J$ is the cost, $\Omega_t$ is the time period of interest, $u$ is the shaping coil power supply voltages, and $e$ is the vector of isoflux shaping errors which can include signals like coil currents, shape control point flux errors, field at the x-points, and the boundary flux or loop voltage. $W$ are weighting matrices for each cost term. In the implementation, the cost function is discretized to discrete time points and equilibria. Note that there are quadratic penalties on the signal values, as well as on the signal first and second derivatives (with respect to time). First-derivative weights are used to penalize trajectories that ramp too aggressively, and second-derivate weights are used to penalize non-smooth trajectories. 

The cost function subject to the FBEE governing equations can be reformulated into a standard-form quadratic program (QP) that can be solved by software packages such as MATLAB's \textbf{quadprog}. This is done by building a prediction model to determine the relationship between actuator commands at every time step and their impact on output errors at every time step. The prediction model technique is similar in principle to the technique used in Model Predictive Control (MPC). Note that unlike traditional MPC, GSPulse performs open-loop optimization; it does not operate in a simulation-type environment where a different QP is solved at every time step. An additional difference is the nonlinear terms of the FBEE model are not linearized, but instead treated exactly and updated at each iteration.

The current implementation of GSPulse is able to solve the cost function for parameters that are linear with respect to voltage or current (electric fields, magnetic fields, shape flux errrors, voltages, and currents) and future work will aim to extend this to additionally optimize for parameters that are nonlinear with respect to currents, such as structural forces. 

The FBEE governing equations are (using a COCOS-17 coordinate system \cite{Sauter2013}):

\begin{equation}
    \Delta^* \psi = -\mu_0 r^2 P' - FF' - \mu_0 r J_\phi^{ext} 
\end{equation}

\begin{equation}\label{eq:circuit}
    v_i = R_i I_i + M_{ij} \dot I_j + \dot \Phi_i^{pla} 
\end{equation}

\begin{equation}
    V_p = -\dot \psi_{bry} = R_p I_p + \frac{1}{I_p}  \frac{d}{dt} \left(\frac{1}{2}L_I I_p^2 \right)
\end{equation}

See \cref{subsec:model_equations} for the definitions of each variable. The first equation is the well-known Grad-Shafranov equation \cite{Grad1958,Shafranov1966} for axisymmetric equilibrium force balance, the second describes axisymmetric circuit dynamics within a tokamak, and the last describes the plasma current evolution. The plasma current governing equation is sometimes presented in different forms. The form here used by GSPulse is the scalar Ejima equation \cite{Ejima1982} which describes how the plasma surface voltage must compensate for plasma resistance and changes to the plasma current and inductance. Note that Ejima equation (and GSPulse) do not consider non-inductive current, but could be extended be replacing the resistive voltage term above with $R_p (I_p - I_{non-inductive})$ \cite{Romero2010}. GSPulse can also be configured to accept the loop voltage or boundary flux directly, instead of computing it from the Ejima model.

The employed solution strategy is to use a Picard iteration scheme for the Grad-Shafranov plasma flux distribution and couple this with a dynamic optimization. This is analogous to the Picard iteration scheme employed in many static equilibrium solvers, that alternates between updating the conductor currents and the Grad-Shafranov equilibrium. The major difference is that in the GSPulse case we include the dynamics as a constraint in the optimization and also solve for all of the voltages and currents over the time window of interest. \Cref{fig:gspulse_convergence} gives a cartoon of the convergence strategy used in GSPulse, where we alternate between the open-loop trajectory optimization and plasma Picard updates of the flux distribution. 

A rough outline of the algorithm is provided below. For a detailed mathematical treatment, see \cref{app:gspulse_algorithm}.

\begin{enumerate}[itemsep=0.5em]

\item \textbf{Configuration:} In this step the user defines a standard set of inputs which includes the tokamak geometry, mutual inductance matrices, coil and vessel resistances, core plasma properties needed to solve the Grad-Shafranov equation such as the P' and FF' profile shapes, constraints on the plasma properties such as the the plasma current, thermal energy, or internal inductance, and the target shapes and weights. Note that GSPulse itself does not derive mutual inductance and Green's function tables from the tokamak geometry, but instead has been configured to accept geometry and inductance tables computed from both the TokSys \cite{Humphreys2007} and MEQ \cite{epflSuite} suites of equilibrium codes. 

For the inductive plasma current drive model, the user can either specify a waveform for plasma resistance, which is used in the Ejima equation, or directly specify the target boundary flux evolution. When using Ejima mode, the plasma internal inductance $L_I := \int_{\Omega_p} B_p^2 / (\mu_0 I_p^2)dV$ is computed via numerical integration of the equilibrium and updated each iteration. Shape targets can be created with the shape editor GUI (\cref{fig:shape_editor}). Weights are specified for various shaping parameters and coil current and coil voltage characteristics. These are used to tune the solution to give a desired trajectory -- for example they can be used to tune how important achieving a particular strike point is, or to penalize non-smooth trajectories for the currents. 

As part of the configuration, the user can also specify the initial conditions for the solver such as the initial coil and vessel currents and initial voltages. If unspecified, the initial conditions are treated as free parameters for the optimization. 

\begin{figure}[H] 
    \begin{center}
        \frame{\includegraphics[width=14cm]{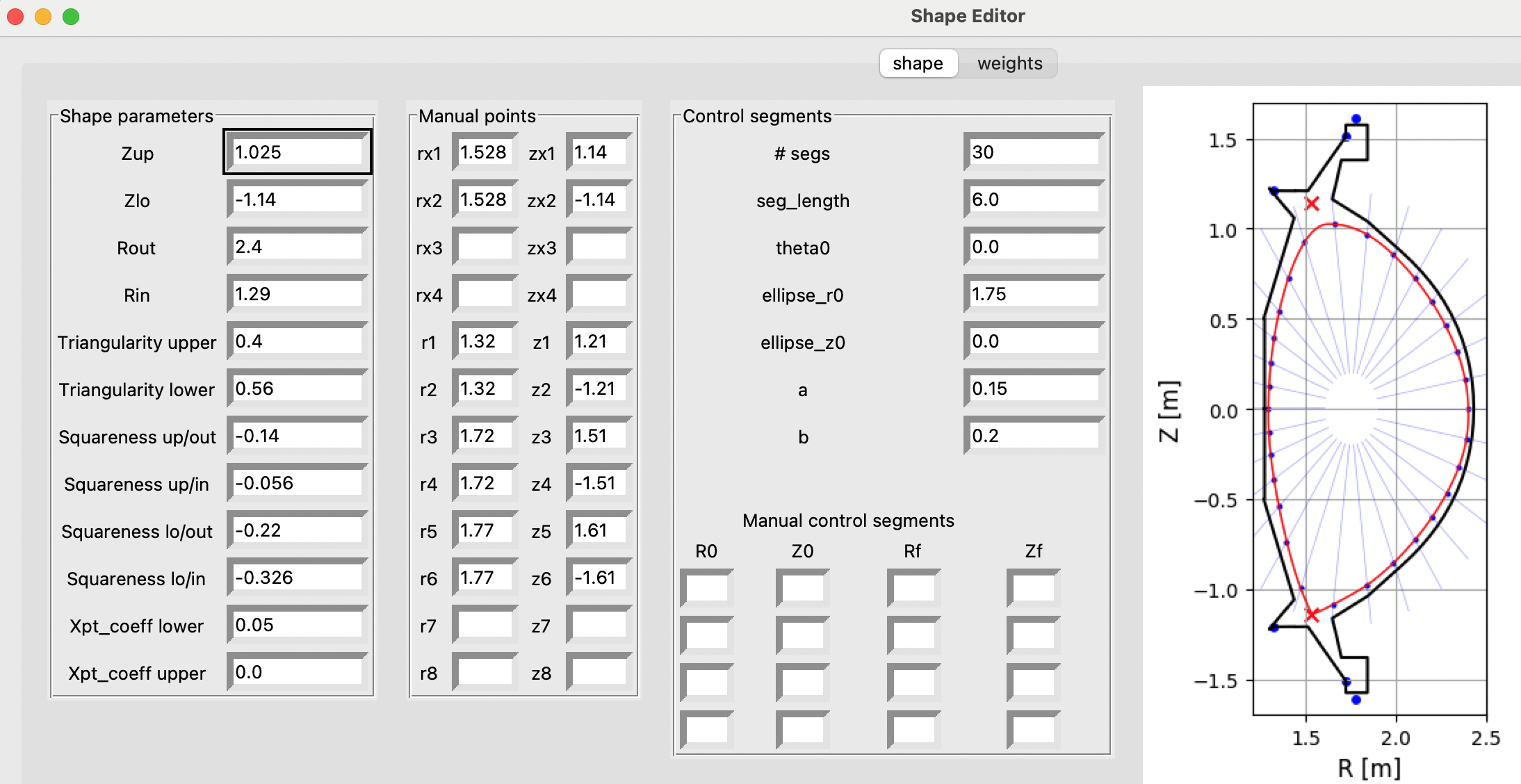}}
    \end{center}
    \caption{GSPulse shape editor GUI using an outline of the SPARC limiter. The user specifies target shapes in terms of a smaller number of parameters such as elongation and triangularity, and the corresponding timebase for the shapes.}
    \label{fig:shape_editor}
\end{figure}

\item \textbf{Initialization:} In this step, based on the target shape and target $I_p$ the algorithm obtains a rough estimate of the plasma current distribution at each step. 

\item \textbf{Compute target boundary flux:} In this step, GSPulse integrates the scalar Ejima equation (\cref{eq:ejima} to compute the boundary flux evolution corresponding to the target $I_p$ evolution. The boundary flux becomes a target for the conductor optimization. 

\item \textbf{Optimize conductor evolution:} Re-formulate the circuit dynamics and shaping cost function in terms of a quadratic program, and solve this QP to obtain the conductor evolution that minimizes shaping errors. 

\item \textbf{Grad-Shafranov Picard iteration:} Given the conductor currents and profiles from the $k^{th}$ iteration, perform a Picard update of the Grad-Shafranov equation to find the flux distribution at the $(k+1)^{th}$ iteration. GSPulse has the option of performing this step natively, or via calling subroutines of the MEQ suite \cite{Moret2015} of equilibrium codes.

\item \textbf{Repeat from Step 3 until convergence.}
\end{enumerate}

\subsection{Vertical stabilization during convergence iterations}\label{sec:vs_convergence}

One consideration is that equilibrium solvers that use Picard iterations can be vertically unstable. It is well-known that, by itself, the Grad-Shafranov Picard iteration

\begin{equation}
    \Delta^*{\psi^{k+1}} = -\mu_0 RP'(\psi^k) - FF'(\psi^k),
\end{equation}

with a free-boundary edge condition is unstable when applied repeatedly without feedback \cite{Johnson1979}. However, this tends to be more of a problem for reconstruction solvers and forward solvers. For example, RT-EFIT \cite{Ferron1998} and the original LIUQE \cite{Moret2015} explicitly add a vertical shift free parameter that could be updated across iterations, and forward simulators like FreeGSNKE \cite{Amorisco2024}, FGE \cite{carpanese2021}, and NICE forward mode \cite{Faugeras2020} avoid Picard iterations in favor of Newton or Newton-like methods.

However, as an inverse solver, GSPulse tends to be robust against this vertical instability. This is linked to the user setting a shape target objective within the optimization, which applies corrective action between iterations to keep the plasma shape from drifting. The stability is also related to the time resolution of the GSPulse configuration. Smaller time steps tend to be more difficult to stabilize. The reasoning is that the numerical (unphysical) instability of the equilibrium Picard iteration does not depend on the time resolution, while it must be counteracted by currents in the coils and conducting structures. These do depend on the time resolution, as the circuit dynamic model (\cref{eq:circuit}) constrains the currents such that $\Delta I \sim M^{-1} v \Delta t$, that is, the amount of compensating current that can be provided decreases with the time step size. When very small time steps are used, at or below the vertical instability timescale, then the vertical instability can reappear and convergence can be sensitive to the exact weights and tuning parameters. 

\begin{figure}[H]
    \begin{center}
        \includegraphics[width=0.6\linewidth]{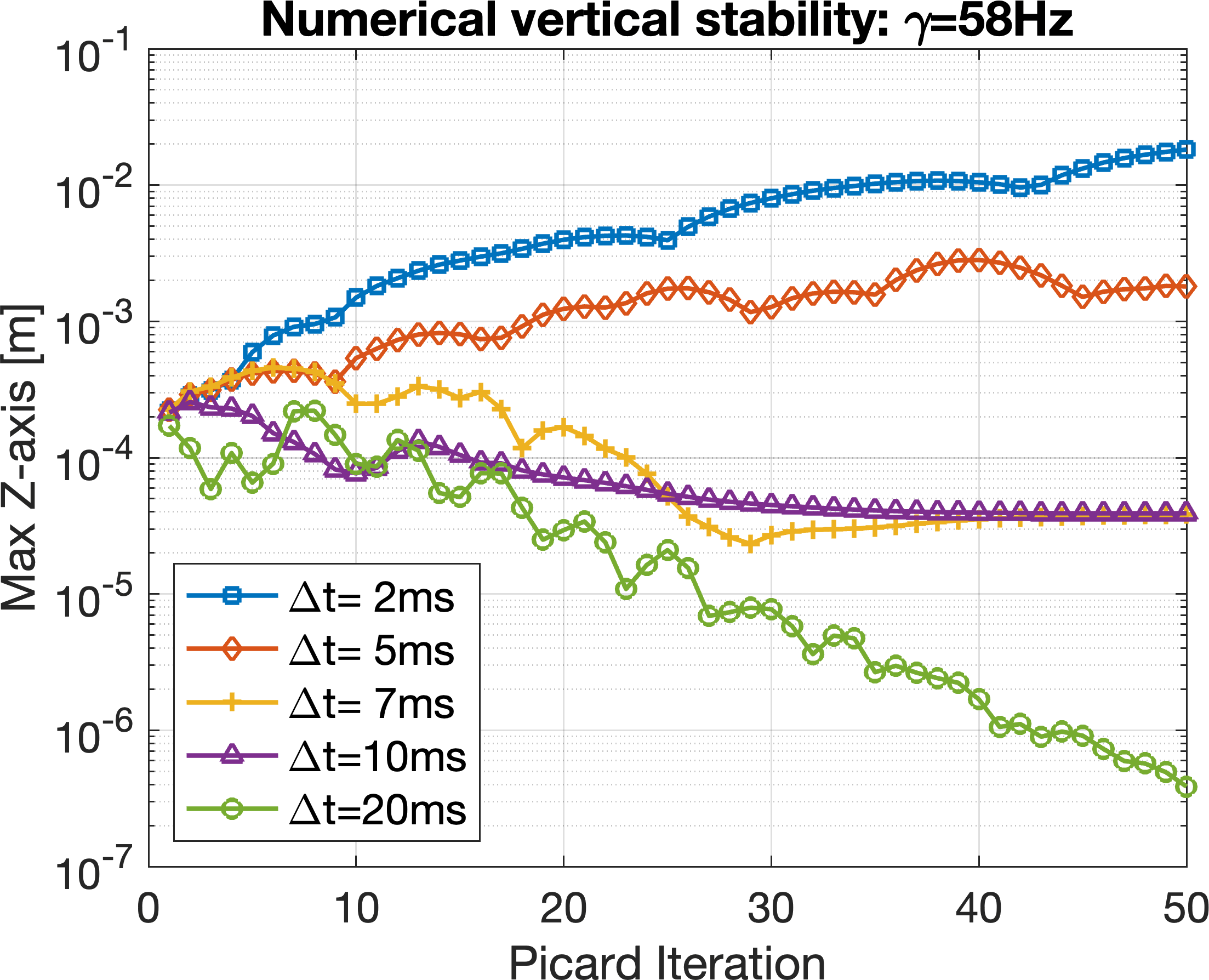}
    \end{center}
    \caption{Vertical convergence as a function of time resolution. The objective is to design 100ms of SPARC equilibria in a balanced up-down double-null configuration. The equilibria are computed to have a vertical growth rate of $\gamma=58$Hz ($\tau =17$ms). With time resolution much smaller than this the algorithm goes vertically unstable, while the largest time resolution of $\Delta t=20$ms shows exponential convergence of the magnetic axis Z-position towards zero.}
    \label{fig:vs_convergence}
\end{figure}

This is illustrated in \cref{fig:vs_convergence}. In this example, we use GSPulse to design 100ms of a flattop Ip=8.7MA pulse for SPARC (\cite{Creely2020}) and maintain an up-down symmetric double null equilibrium. The vertical instabilty growth rate of this equilibrium is computed via MEQ to be $\gamma=58$Hz ($\tau =17$ms). The time resolution for this pulse is varied from 2-20ms (between 6 and 51 equilibria each) and the maximum value of the magnetic axis Z-position across the sequence of equilibria is plotted. The largest time step of 20ms has exponential convergence in driving the z-axis to exactly zero. At intermediate time steps, the sequence converges to a small sub-mm fixed value, while the 2ms and 5ms resolution pulses go vertically unstable. The exact conditions for whether the vertical stability is developed or avoided depend on interactions between the plasma equilibrium, the time resolution, and the shaping tuning weights, and a detailed study is beyond the scope of this article. However, we note that in typical pulse planning applications it is often desired to evaluate larger time steps over long time horizons, conditions over which the algorithm is more robust against vertical instability leading to few issues in practice.

\subsection{Weight design and specification}

An important element of achieving a specific pulse trajectory is configuring the weights used by the optimizer. GSPulse has been implemented to optimize for a number of signals related to voltage, current, and isoflux error measurements, that is, the flux or field at specified locations or linear combinations thereof. Isoflux-based signals are used, in contrast to spatial-based signals (like the shape control gaps), because this results in linearizations that are not equilibrium-dependent. It is discussed in more detail in \cref{app:gspulse_algorithm}.2 (step 3) that there is a computational advantage to this approach which allows for large matrix calculations to be re-used. 

The error signals currently supported are: power supply voltage, coil currents, linear combinations of coil currents (for example, to penalize up-down asymmetry in coil currents), the absolute flux at specified locations, the relative flux at specified locations (that is, the difference between flux measured at point A vs point B, which can be used for example to ensure that the flux at specified shape control points is equal to the flux at a desired touch point or x-point), and the radial and vertical field at specified locations (for example, to form an x-point by driving the field at a specified location to zero). 

The cost associated with any single error signal, at a single time step is: 

\begin{equation}
    J = e^2 w_e + \dot e^2 w_{\dot e} + \ddot e^2 w_{\ddot e}
\end{equation}

A useful heuristic is to design the weights such that cost for each error signal is of order unity. This is achieved by specifying the weights such that they are inversely proportional to the square of typical error range values:

\begin{equation}
    w_e \sim \frac{1}{e_{typical}^2} ,\;\;\;\;   w_{\dot e} \sim \frac{1}{\dot e_{typical}^2} ,\;\;\;\;  w_{\ddot e} \sim \frac{1}{\ddot e_{typical}^2}     
\end{equation}

To illustrate, consider the following typical pulse scenario objective (the example in \cref{subsec:sparc_divert} approximately matches this): the plasma begins limited, and then transitions to a lower single null, diverted phase. The shape should be controlled relatively accurately, meaning that large scale features like plasma volume, elongation, and triangularity should be actualized. However, it is also desired to minimize coil voltages and currents and therefore some tradeoff in shape is allowed to reduce actuator usage. 

To achieve this, one could design the weights as follows (the description of these weights are modified from the \cref{subsec:sparc_divert} example, which has a slightly different objective): 

\begin{itemize}

    \item To control the shape while it is limited, we assign weights to penalize the difference in flux between the shape control points and the touch point location. When it is diverted, we penalize the difference between shape control points and the target x-point location. The weights are specified as a function of time and we smoothly vary between these sets of weights at the limited-diverted transition time. 

    On SPARC, a 1cm spatial shape control gap error corresponds to 0.01Wb (near the x-points) up to 0.4Wb (at the outer midplane). Applying the inverse-square heuristic this corresponds to a weight of $w_e = 10^2 - 10^4 \text{ per Wb}^2$. For simplicity, we assign a weight of $10^3/\text{Wb}^2$ to all control points. 

    \item To form the x-points, we assign a target of $B_r=B_z=0$ at the desired locations. We would like to form these with approximately 1cm precision, which for an 8.7MA SPARC equilibrium corresponds to $~0.05$T so the weight is taken as $w_e = 1/(.05\text{T})^2 = 400/\text{T}^2$. The time-varying weight is assigned such that it is zero during most of the limited phase but transitions so that the x-point is formed slightly before the plasma diverts. 

    \item To form a lower single null equilibrium we control the relative flux difference between the upper and lower x-points and assign it a non-zero target value. Since 1cm flux difference at the outer midplane corresponds to 0.4Wb, the target flux difference is assigned such that $\Delta r_{sep} = 5\text{mm} \implies 0.2\text{Wb}$. The weight is assigned equal to the other control point flux weights. 

    \item To control the surface voltage, we penalize the time derivative of the absolute value of the flux at the plasma boundary because $V_p = -\dot \psi_{bry}$. From the Ejima equation we have that $V_p \sim L_I \dot I_p / I_p$ and on SPARC $L_I$ is approximately unity. Taking the desired precision to be $10$kA/s at $I_p=8.7$MA results in a typical $\dot \psi_{bry}$ error of $10^-3$Wb/s, resulting in a weight of $w_{\dot e} = 10^6 / \text{V}^2$. 

    \item The inverse-square heuristic is also applied to the coil currents and voltages, resulting in weights for these parameters of $10^{-6}/\text{A}^2$ and $10^{-4}/\text{V}^2$. Additionally, to find smooth trajectories, we apply a weight on the second time derivative of the coil currents. The weight is initialized at a small value and increased as much as possible until it is observed to affect the shape no more than several mm, with a final value of $w_{\ddot e} = 10^{-9}\text{(A/s$^2$})^2$. 

\end{itemize}

\section{Results}

\subsection{Validation against NSTX-U 204660}

\begin{figure}[H] 
    \begin{center}
        \includegraphics[width=0.9\linewidth]{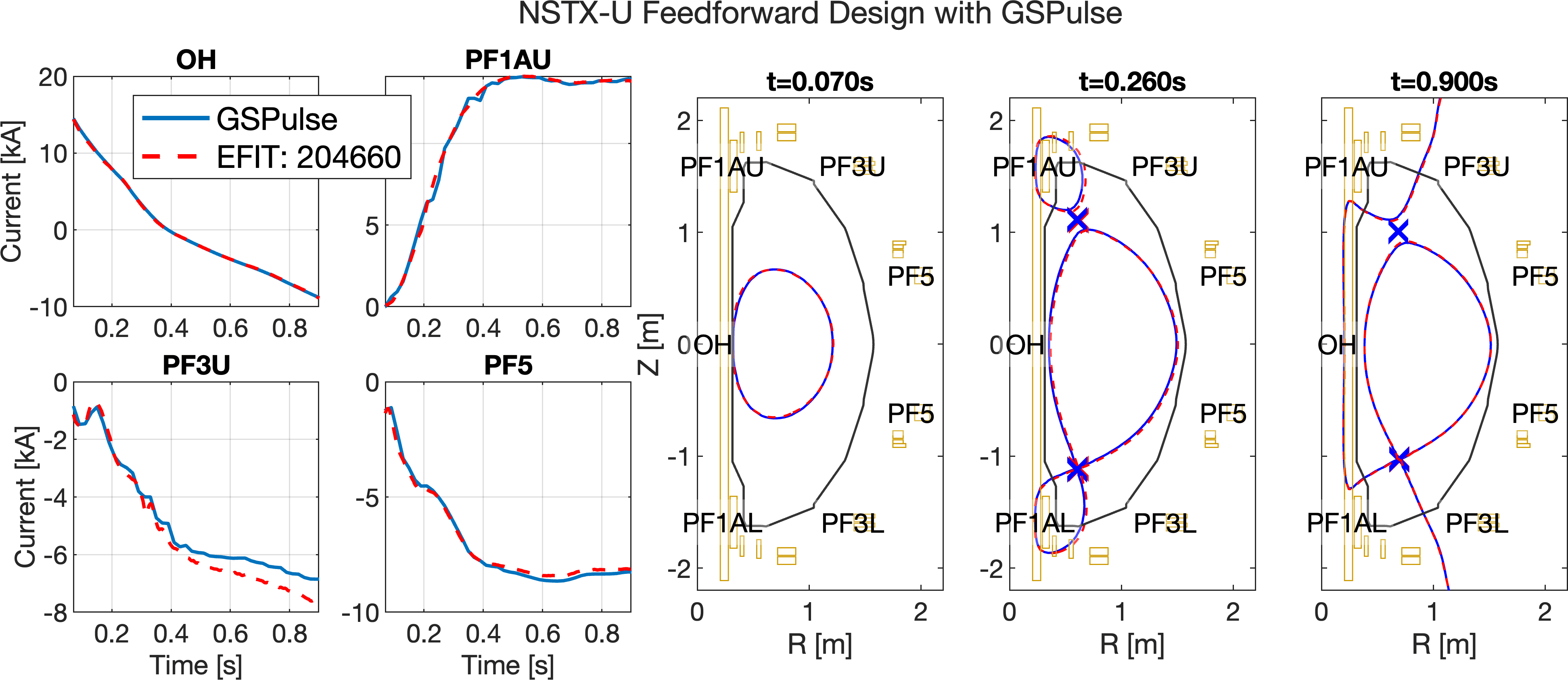}
    \end{center}
    \caption{Validation results of GSPulse against NSTX-U shot 204660. The coil currents are obtained from EFIT01 (magnetics-only). Flattop for this pulse is at 0.38 seconds but the non-uniform field produced by the OH coil requires the PF3 current to continue drifting during flattop in order to maintain core plasma shape. The result is that the outer strike point is not held constant during the pulse if the core shape is also fixed.}
    \label{fig:nstxu_gsulse}
\end{figure}

GSPulse has been validated against experimental results and simulations for several tokamaks including NSTX-U \cite{Berkery2024} and MAST-U \cite{Harrison2019}. \Cref{fig:nstxu_gsulse} illustrates a validation study against NSTX-U shot 204660. The objective of this GSPulse run is to re-create the exact progression of equilibria and coil currents that was achieved during the actual feedback-controlled pulse \cite{Boyer2018}, as defined by the NSTX-U EFIT01  reconstruction \cite{Sabbagh2001}. Since GSPulse internally enforces the FBEE model, we expect that any modeling discrepancies or implementation errors would prevent the algorithm from being able to exactly match both the target shapes and coil currents, up to the limit of accuracy for EFIT01. 

For this validation case, GSPulse is configured as follows:

\begin{itemize}
    \item The initial condition (equilibrium, coil currents, and vessel currents) are provided from an EFIT \cite{Lao1985} reconstruction at $t=50$ms. The  evolution of the boundary flux is specified as a direct target based on the EFIT01 (magnetics-only) loop voltage. Core plasma properties (stored thermal energy, $I_p$, and P' and FF' profile basis function shapes) are also obtained from the EFIT01 values. 

    \item The shape targets are derived from the actual shape progression of the pulse. This is a non-optimal shape progression since it contains disturbances and feedback oscillations, but is used for the validation case. For shape targets, we specify only the x-point locations and the core plasma boundary. The strike points are not included as targets in this case. 

   \item The coil current trajectories from shot 204660 are included as targets for GSPulse and weighted lightly. Of course the advantage of GSPulse is that it can identify new, optimal coil trajectories and in general it is not necessary to a-priori specify coil currents. However, they are included for the validation run in order to guide the solution to the best experimental fit, since some coils can have similar impacts on the equilibrium.
       
    \item We solve for an equilibrium at every 20ms interval between t=50ms and t=910ms (44 equilibria in total). This calculation takes 1-2 minutes to run on a laptop. 
\end{itemize}

The results of the validation run are shown in \cref{fig:nstxu_gsulse}. In general we see robust agreement for both the achieved equilibria and current trajectories. The current in the PF3U coil begins to deviate from the experimental values in the second half of the pulse, reaching a variation of -7.2kA versus -8kA by end-of-pulse. This hints at a small amount of model discrepancy; for example, inaccurate estimates of the distribution of resistances and inductances within the vessel wall could have this type of effect. It is also possible that this discrepancy could be resolved with more refined estimates of the internal profile evolution. Since EFIT01 uses simple low-order polynomials for the internal profile shapes, and the GSPulse validation cases enforces the same profile evolution, inaccurate EFIT01 profiles could lead GSPulse to inaccurate estimates of the induced vessel currents. 

One of the emergent features from this run, observable in the equilibria of \cref{fig:nstxu_gsulse}, is that the outer strike point moves radially outwards throughout the pulse, even though the inner strike point position, x-point position, and core boundary shape remain relatively fixed. This NSTX-U pulse used a subset of the divertor shaping coils and consequently could not control both the outer strike point and x-point positions. The motion of the strike point is driven by a change in the time-varying current of the OH coil which does not supply a uniform field, a known feature of the NSTX-U design \cite{Menard2012}. Since GSPulse is adequately capable of modeling this type of effect, it could be a useful operator tool for evaluating the tradeoffs in shape evolution for different choices of coils and coil limits under time-varying conditions. 

\subsection{Validation against MAST-U 48092}

\begin{figure}[H]
    \centering
    \includegraphics[width=0.9\linewidth]{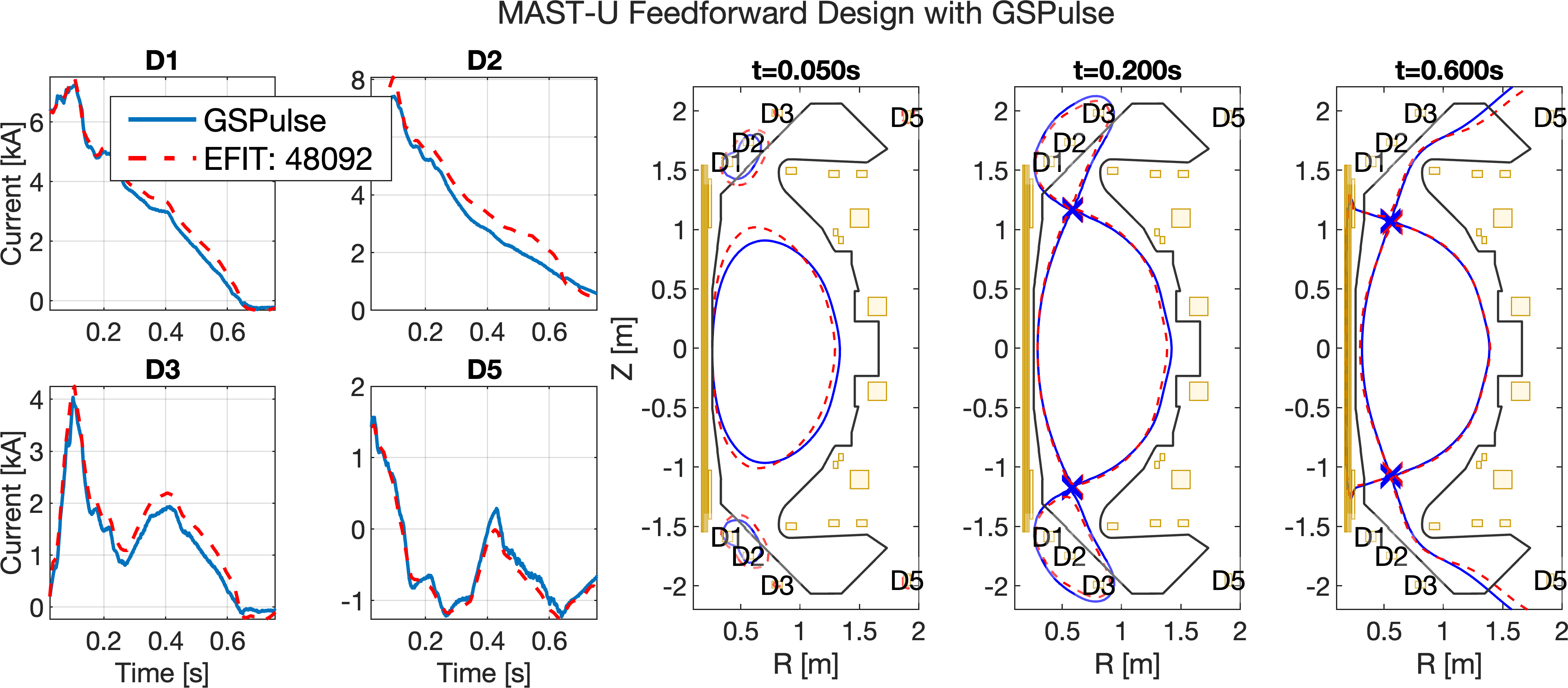}
    \caption{Validating GSPulse against MAST-U shot 48092 shows consistency between predictions and experiment.}
    \label{fig:mastu_gspulse}
\end{figure}

\Cref{fig:mastu_gspulse} illustrates a similar validation calculation performed for MAST-U shot 48092. The configuration setup is nearly identical to the NSTX-U validation case above. In particular, the initial condition and plasma properties are taken from the EFIT reconstructions. Large weights are applied for matching the target shape evolution, and small weights are applied for matching the coil currents. Again, the goal is to find a GSPulse solution that best-matches the experiment, and verify that the internal FBEE model is consistent with reality. The run is set up to solve for equilibria every 10ms between t=25ms and t=755ms. 

The results for this run are shown in \cref{fig:mastu_gspulse} again showing strong agreement in the coil currents and equilibrium progression. There are some slight variations in the currents, and the early plasma boundary is slightly less elongated (order 5-10cm). This is reasonable, as during the earliest portions of the pulse, the combination of low plasma current and large vessel currents make the equilibrium more sensitive to the wall model inductances and resistances. During the pulse, the strike leg is moved radially outwards into the super-X divertor chamber. There is a 5cm discrepancy in the strike leg position which also hints at some model mismatch in this area of the tokamak. 

For these validation cases, the configuration has been setup to exactly match the actual pulse. However, after model verification, it is straightforward to set up GSPulse to identify desirable trajectories that are not a-priori specified. Future work will also aim to improve the trajectories beyond the existing scenarios, for example, finding smoother and less aggressive coil waveforms, reducing the amount of time that the plasma is limited, and bringing the strike point or super x-points into position faster. 

\subsection{Design of SPARC strikesweep scenario and validation against FGE}

We also validate GSPulse against the FGE \cite{carpanese2021} code during a strike point sweeping simulation for the SPARC tokamak. FGE has been used and tested extensively on a number of tokamaks. It was also the simulation environment for a reinforcement learning magnetic controller \cite{Degrave2022} that controlled the TCV tokamak, demonstrating FGE's validity for predictive performance.

\begin{figure}[H]
\includegraphics[height=7cm]{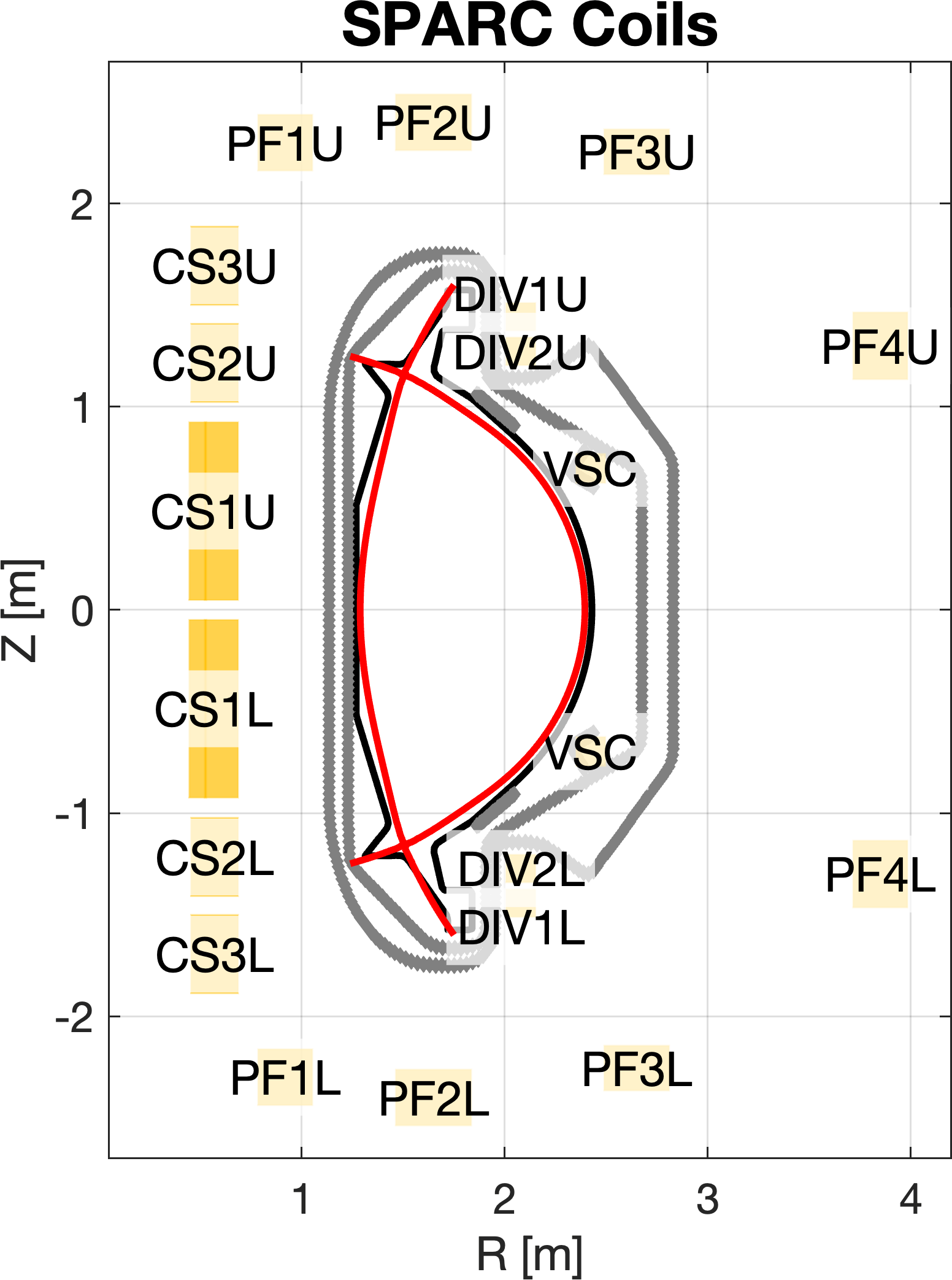}\hspace{0.5cm}
\includegraphics[height=7cm]{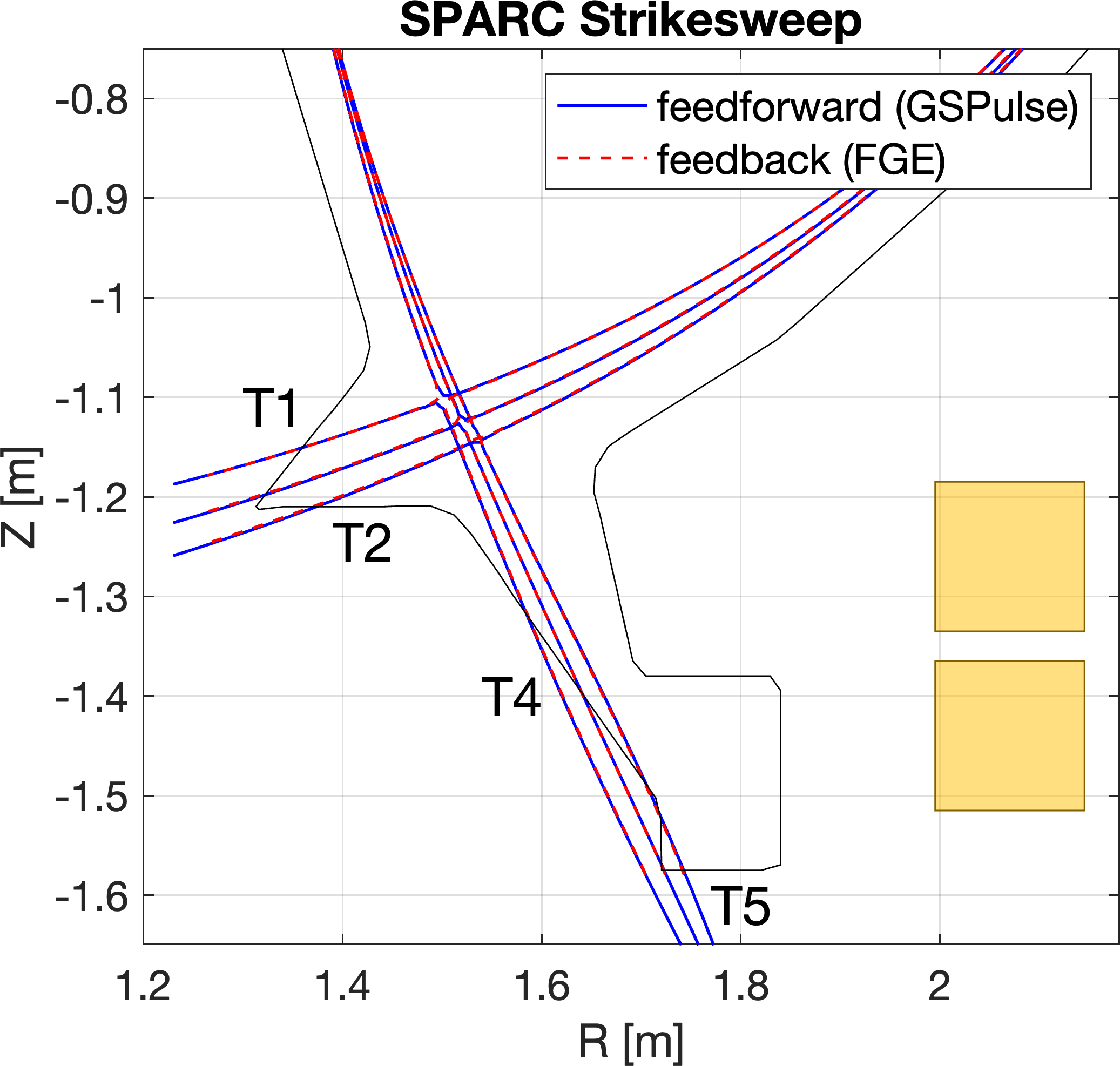}\\
\caption{\textbf{a)} Geometry layout of the SPARC shaping coils. \textbf{b)} The equilibrium last-closed flux surface at different times within the strike point sweeping scenario. The feedforward results are obtained with GSPulse and show good agreement with the full closed-loop feedback control simulation computed with FGE. Note that this example is an early exploratory strikesweep scenario, and the SPARC design has evolved to disallow sweeping the inner strike point across the Tile 1/ Tile 2 boundary.}
\label{fig:sparc_sweep}
\end{figure}

This validation case is configured as follows: 

\begin{itemize}
    \item The initial condition is an 8.7MA, double-null, equilibrium. GSPulse is used to design the coil current and voltage trajectories for sweeping the strike points of this equilibrium across the divertor tile surfaces, as shown in \cref{fig:sparc_sweep}b, at a frequency of 0.8Hz. 

    \item The FGE feedback simulation is configured to use the same initial condition as GSPulse. Since FGE is a pulse simulator, we need to specify vertical feedback control within the loop. A feedback controller is designed using the VSC (Vertical Stability Coil) as the primary actuator. The results of the GSPulse computation are used as feedforward inputs to the simulation, so that the output voltages supplied to FGE at each simulation time step are a sum of the shape feedforward and vertical feedback controller commands.
    
    \item Both FGE and GSPulse use the same geometry definitions, inductance matrices, and resistances. GSPulse is configured to solve for the voltages and equilibria every 50ms, and FGE is configured for a time step of 1ms. 
    
\end{itemize}

\begin{figure}[H] 
    \begin{center}
        \includegraphics[width=15cm]{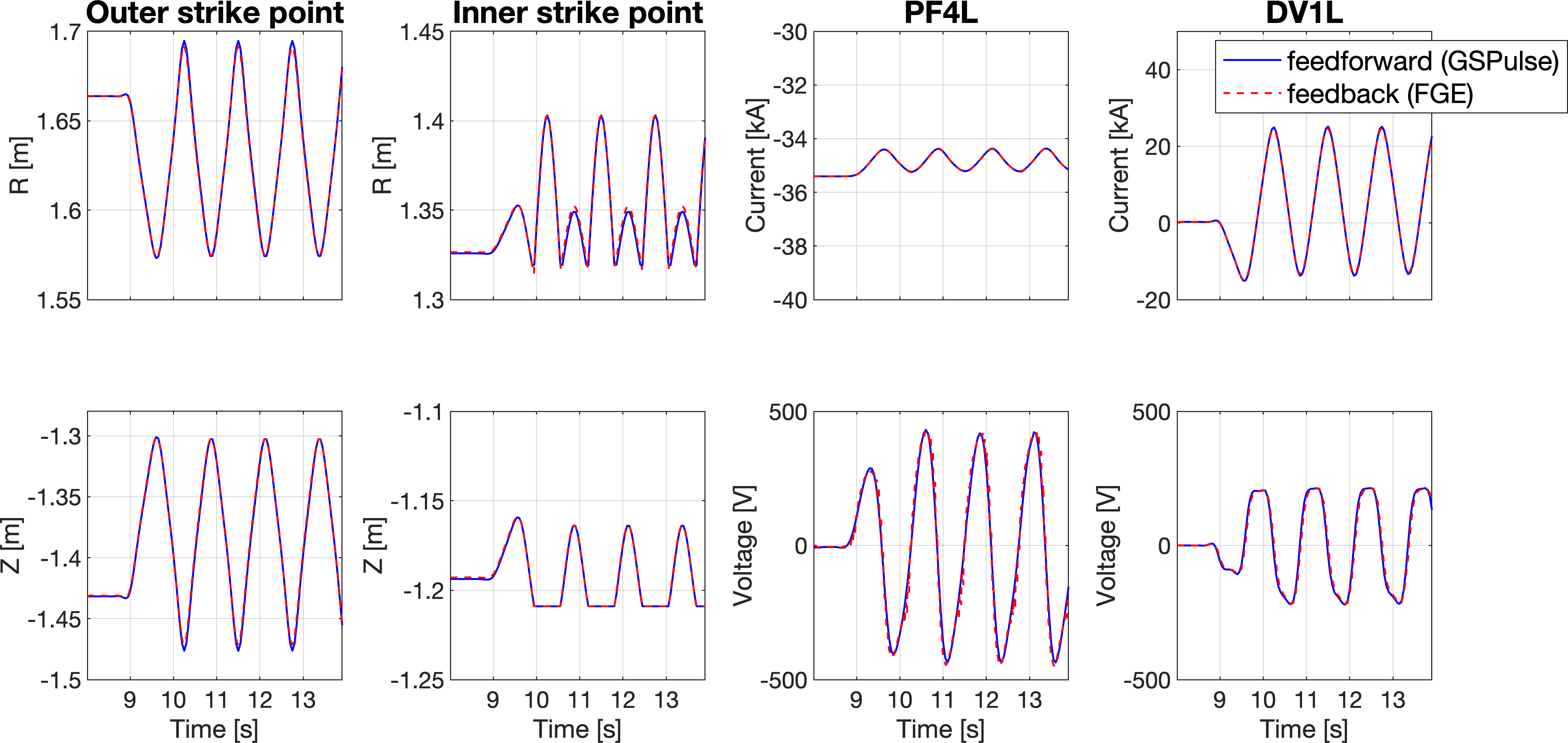}
    \end{center}
    \caption{Representative time traces for the strike point sweeping scenario, comparing the feedforward predictions from GSPulse agsinst the full closed-loop feedback control simulation using FGE. The strike point positions agree to within 4mm at all times.}
    \label{fig:sparc_sweep_traces}
\end{figure}

The pulse-planning (GSPulse) and pulse-simulating (FGE) results are shown in \cref{fig:sparc_sweep,fig:sparc_sweep_traces}. There is strong agreement between these models, again illustrating that GSPulse is reliably solving the FBEE problem and can be used predictively. \Cref{fig:sparc_sweep_traces} shows time traces for the strike point locations and representative currents and voltages for 2 of the 19 coils. GSPulse and FGE agree on the strike point positions to within  4mm at all times (note that because of the strike point angle-of-incidence this generally corresponds to $<1$mm precision of the plasma boundary). One interesting point is that the target strikesweep was specified as a triangle wave, but the GSPulse-computed voltage trajectories (for example the DV1L voltage trajectory) can differ significantly from a triangle wave, suggesting a sophisticated input-output relationship that could be difficult to design by hand.

\subsection{SPARC scenario design: plasma breakdown and startup}

\begin{figure}[H] 
    \begin{center}
        \includegraphics[width=8cm]{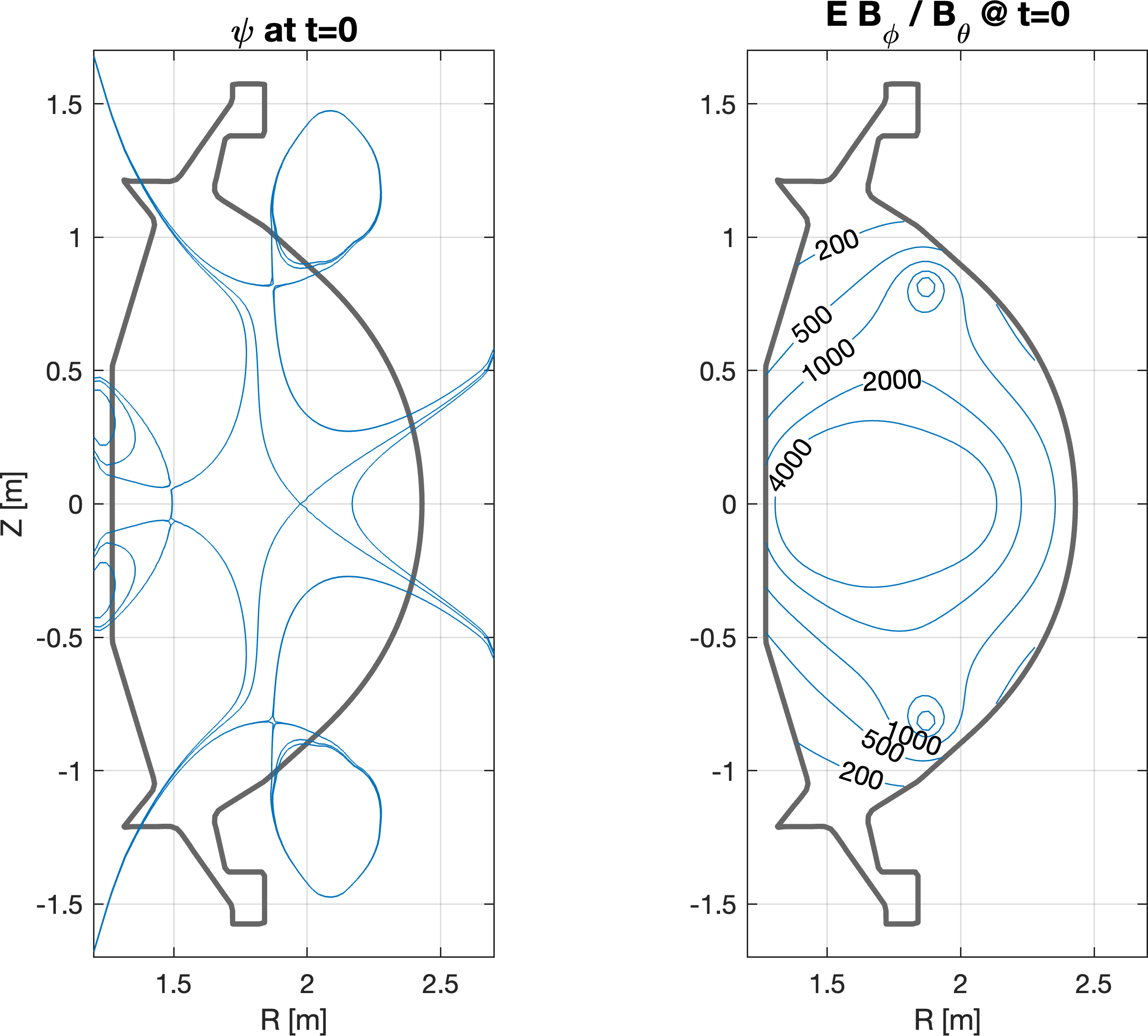}
    \end{center}
    \caption{\textbf{Left:} magnetic flux surfaces at $t=0$ with multiple x-points implying a large area with low $B_p$. \textbf{Right:} Contours of $E_\phi B_\phi / B_\theta$ indicating that there is a large area satisfying the $>1000V/m$ requirement from \cref{eq:breakdown_reqs}.}. 
    \label{fig:breakdown_null}
\end{figure}

\begin{figure}[H] 
    \begin{center}
        \includegraphics[width=8cm]{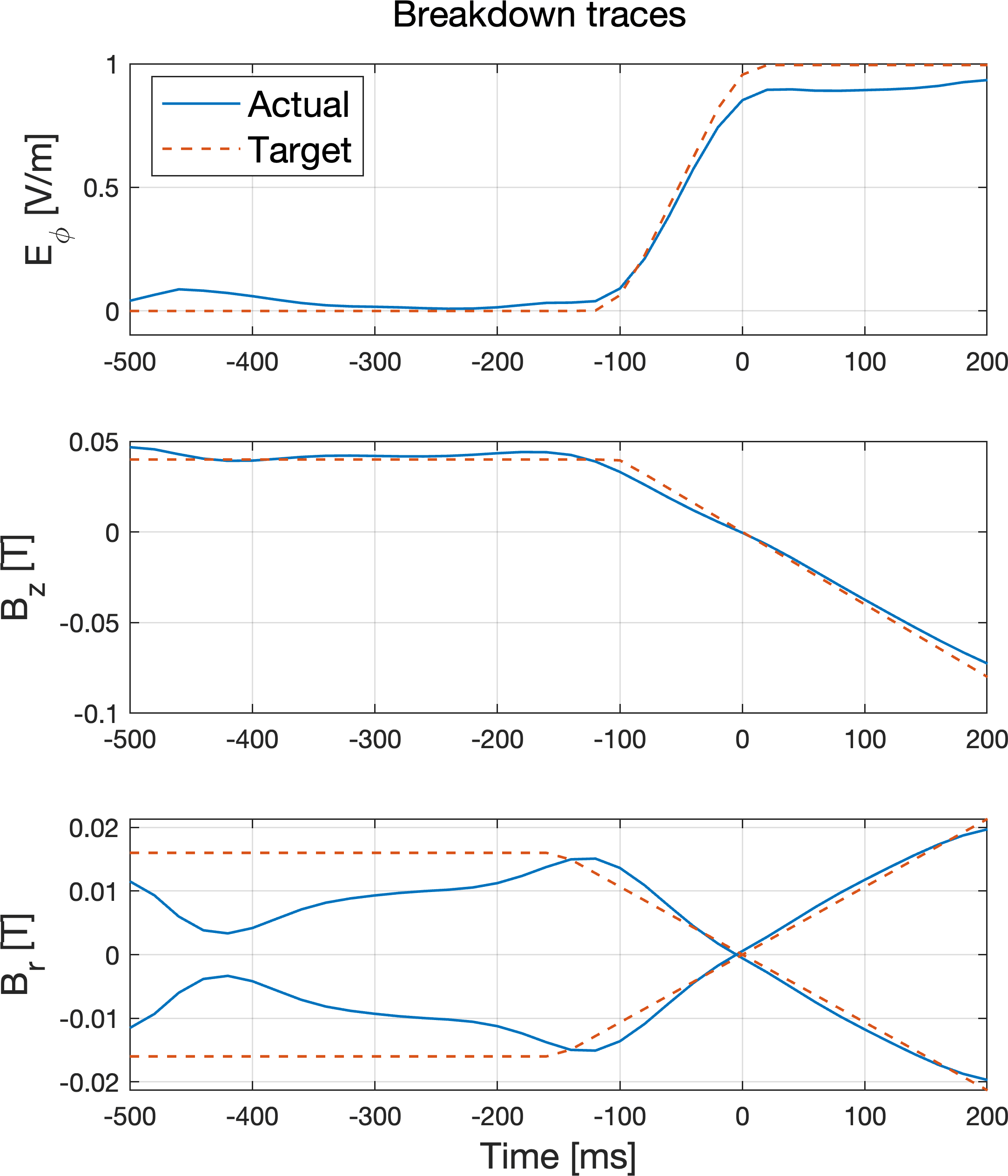}
    \end{center}
    \caption{Target and achieved breakdown traces for electric field (evaluated at R=1.6m, Z=0m), vertical field (evaluated at R=1.6m, Z=0m) and radial field (evaluated at R=1.6m, Z=$\pm$0.6m). The target trajectories were designed following principles from \cite{Leuer2010}, and the achieved trajectories were computed with the GSPulse optimization.}
    \label{fig:breakdown_traces}
\end{figure}

Feedforward is most commonly employed during the early phases of a pulse, when the sequencing for plasma breakdown and startup is critical, and at low $I_p$ before reliable equilibrium reconstructions are available. GSPulse is a natural tool for designing the electromagnetic trajectories required for plasma breakdown since this is a trajectory optimization of the vacuum fields. Similar optimization work has also been performed for other tokamaks in \cite{diGrazia2024,diGrazia2025,Leuer2010}. \Cref{fig:breakdown_null,fig:breakdown_traces} illustrate an example designing plasma breakdown conditions for SPARC. The electromagnetic conditions required for reliable plasma startup can be summarized as \cite{Lloyd1991,Leuer2010}: 

\begin{equation}\label{eq:breakdown_reqs}
\begin{aligned}
    E_\phi &> 0.3 \text{ V/m, no ECH pre-heating} \\
    E_\phi B_\phi / B_\theta &> 1000 \text{ V/m}
\end{aligned}
\end{equation}

These are ``instantaneous'' conditions that must be satisfied near the plasma formation point (typically near the center of the machine or on the inboard limiter) in order to achieve main ion burn-through and plasma breakdown. However, in practice a number of additional time-dependent characteristics are necessary or desirable. These include: avoiding accidentally creating breakdown conditions (reverse breakdown) before the desired start time, sustaining high electric field for a sufficiently long time to burn through the main plasma ion impurities, achieving a vertical field evolution that provides radial force balance as the plasma current channel forms and grows, achieving a radial field evolution that transitions from field null conditions to providing passive vertical stability on a reasonable timescale such that the growing plasma does not go unstable, and pre-charging and balancing the coil currents to maximize the flux that is available for inductive current drive during the pulse. For additional detail on specifying the breakdown trajectory targets, we refer readers to Leuer's work for KSTAR and EAST \cite{Leuer2010}. However, assuming that the target electric and magnetic field evolution have been identified with the appropriate physics modeling, finding the feedforward currents and voltages is easily transformed to a target-tracking optimization problem for GSPulse.

From the GSPulse perspective, the objective can be summarized as:

\begin{itemize}
    \item Track a trajectory for $E_\phi(t) = -\frac{1}{2\pi R_0}\dv{\psi}{t}$.
    \item Track trajectories for $B_r(t), \; B_z(t)$.
    \item Maximize $\psi$ at $t=0$.
    \item Satisfy coil current and voltage constraints.  
\end{itemize}

The location for evaluating $E,B$ and $\psi$ could be either area-averaged quantities usually near the center of the tokamak or at specified locations and is generally an operator preference or tuning parameter. The target trajectories for SPARC at specified locations are shown in \cref{fig:breakdown_traces} and indicate that the breakdown conditions can be reliably met on SPARC. These results are generally consistent with previous breakdown simulations where coil current and voltage trajectories were tuned by hand \cite{Wai2023,Wai2022b}. The peak electric field is 0.9V/m and limited by the power supply voltage capabilities of the CS1 coil. For $B_r(t)$ and $B_z(t)$, the most critical phase is from $t=-100$ms onwards and we only apply high weights during this period, allowing for some tracking error mostly present in $B_r(t)$ before this time. Before $t=-100$ms, the main criterion is to have nonzero field in order to prevent reverse breakdown conditions \cite{Battaglia2019} where a reverse bias of the electric field can cause premature breakdown. Note that this is not strictly required for this scenario where $E(t)$ is approximately zero before breakdown, which is achieved by holding the CS1 current steady at max current, but this is not true for all SPARC breakdown scenarios. GSPulse is able to find a larger area-averaged $E_\phi B_\phi / B_{\theta}$ than was previously found by hand. However, the main advantage is that the operator will be able to tune the breakdown scenario by specifying the desired signal waveforms directly, instead of having to simultaneously adjust and balance multiple coils. 

Currently within GSPulse, the plasma startup optimization (vacuum-only) and standard equilibrium evolution (vacuum optimization plus plasma Picard iterations) are treated as separate problems, due to the numerical difficulty of converging very low $I_p$ equilibria that are present shortly after $t=0$. However, future work will extend GSPulse to solve across this transition period and perform breakdown optimization simultaneously with the equilibrium evolution in the first few seconds of the pulse. 

\subsection{SPARC scenario design: diverting as fast as possible}\label{subsec:sparc_divert}

\begin{figure}[H]
\centering
\includegraphics[width=.3\textwidth]{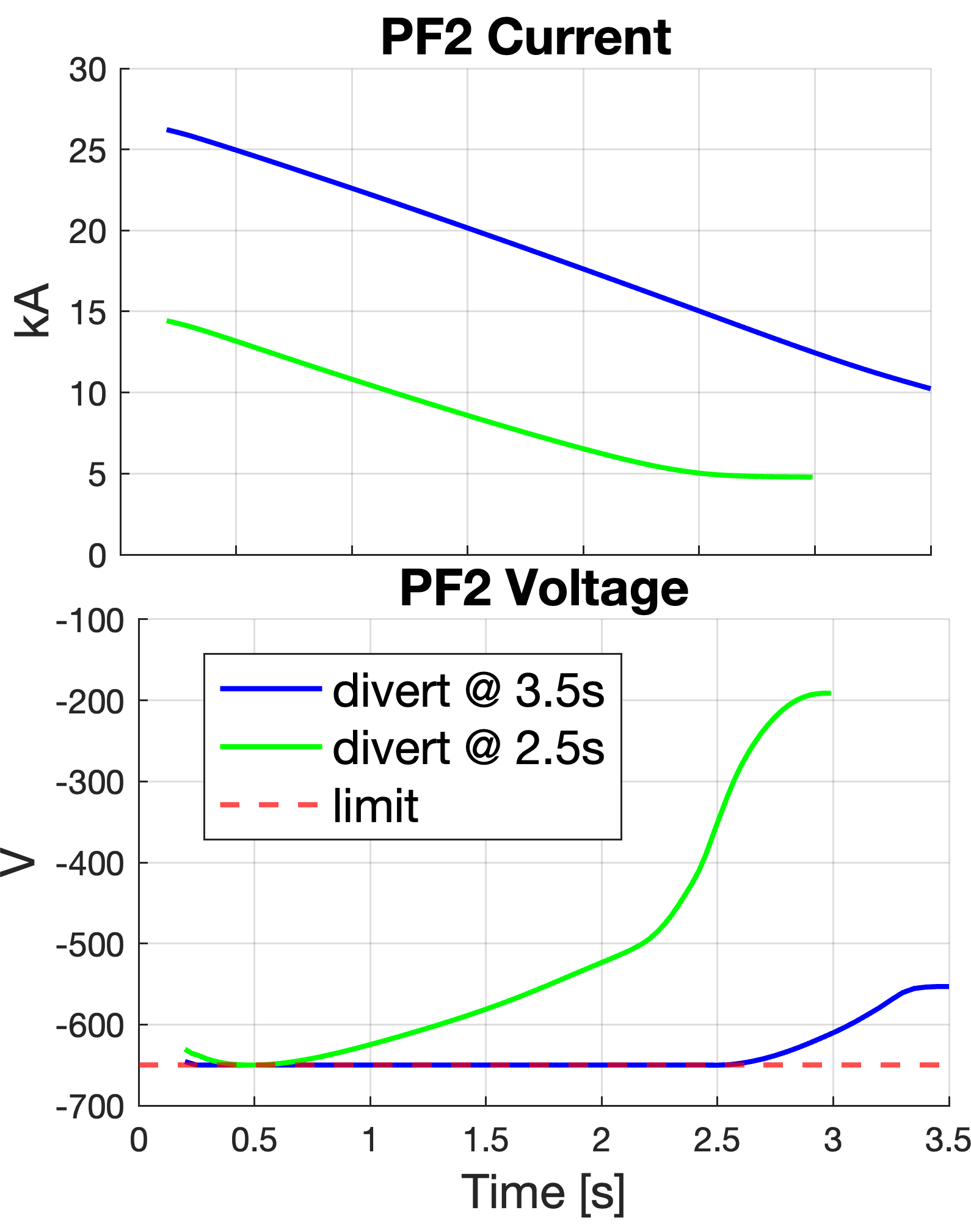}\hfill
\includegraphics[width=.3\textwidth]{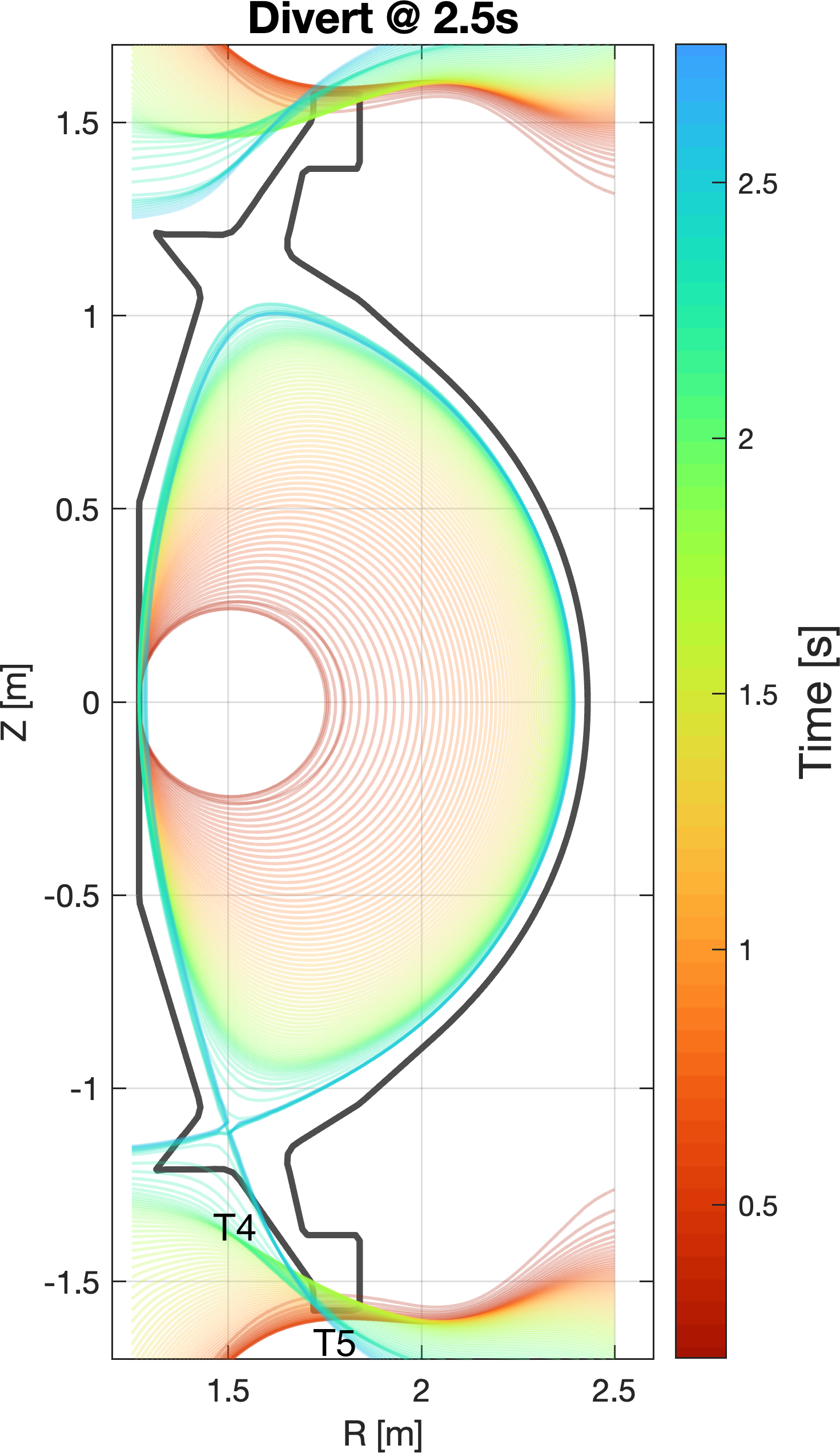}\hfill
\includegraphics[width=.3\textwidth]{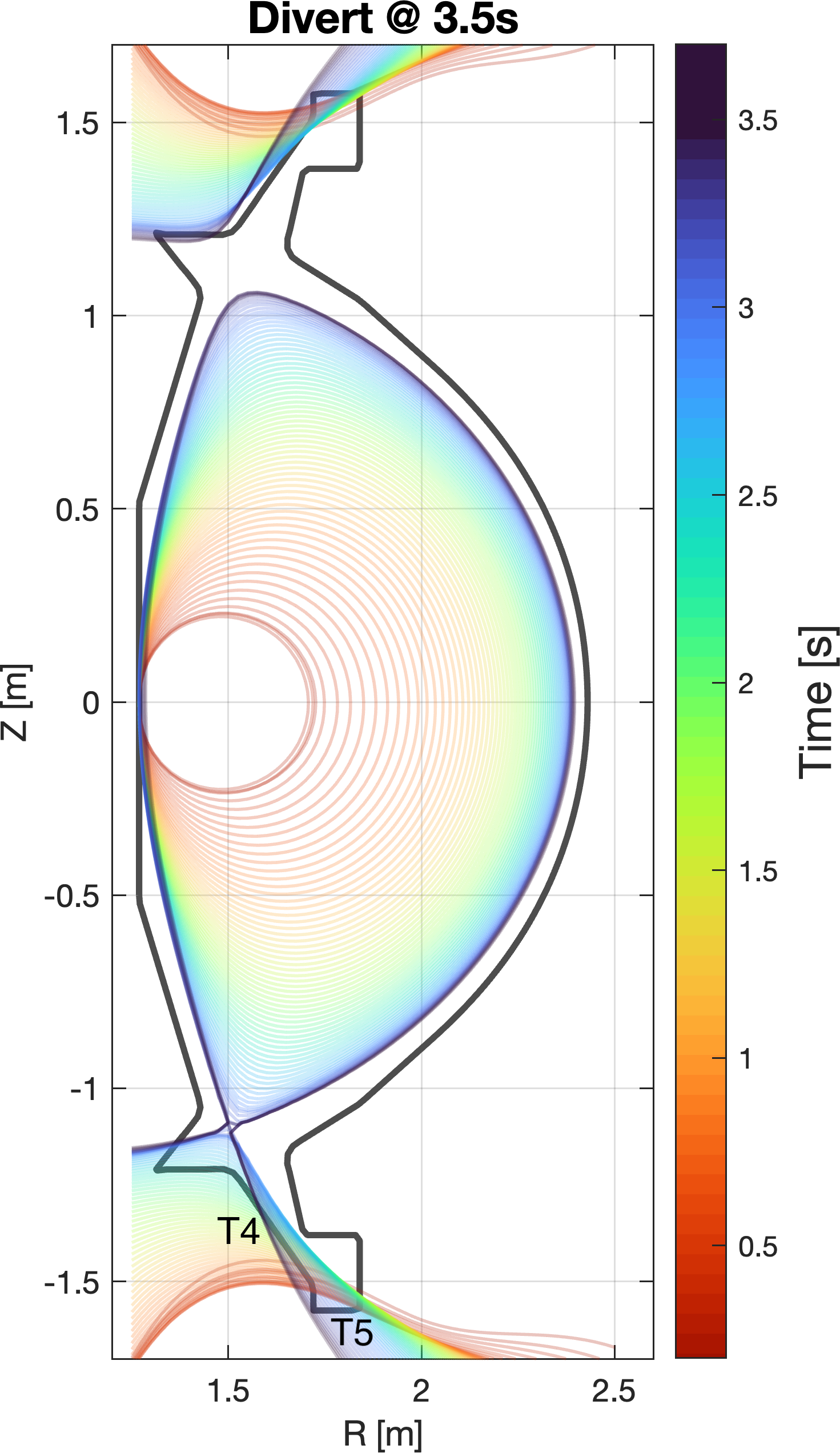}
\caption{Optimizing a SPARC scenario in order to minimize the time required to divert the plasma. With the nominal plasma breakdown scenario, it takes 3.5 seconds to divert because of the requirement to keep the strike point from hitting tile 5 which is limited by the PF2 current evolution. If PF2 current at $t=0$ is allowed to be a free parameter of the optimization then GSPulse identifies a lower starting current that keeps the outer strike point leg straighter and allows for diverting by 2.5 seconds.}
\label{fig:fast_divert}
\end{figure}

\Cref{fig:fast_divert} gives an example of using GSPulse to improve the scenario design for the objective of diverting the plasma as fast as possible. Given an initial $I_p=200$kA equilibrium shortly after plasma breakdown, the goal is to ramp the plasma current at 1MA/s and simultaneously divert the plasma as fast as possible. Diverting is beneficial because it reduces the influx of impurities to the plasma and allows for earlier ICRF heating. Restrictions are imposed that the strike points must be placed only on the divertor tiles, and that the outer strike point avoid contact with tile 5 which is not intended for high heat flux. 

Within GSPulse, this objective is set up as a point A to point B optimization. In other words, the initial condition is fully specified, and the final equilibrium shape is also specified, but there are no shape targets or weights associated with the intermediate times. There are only penalties on the smoothness of the trajectory. We solve for an equilibrium every 30ms or roughly 80-120 equilibria in these scenarios, scanning the endpoint time until the trajectory results are satisfied. 

The results are summarized in \cref{fig:fast_divert}. With the initial condition specified by the nominal breakdown scenario, the time to divert is 3.5 seconds. The challenge in this case is ramping the PF2 coil current fast enough to reduce the curvature on the outer strike point so that it always stays on tile 4 and avoids touching tile 5 (labeled T4/T5 in \cref{fig:fast_divert}). Note that the isoflux line is allowed to touch tile 5 when the plasma is still limited, but not when diverted. This constraint is verified manually and is not part of the GSPulse optimization constraints, except that the target strike point is specified. If we loosen the restriction on the initial plasma current equilibrium, and allow the PF2 current to float as a free parameter, then this reduces the time to divert to 2.5 seconds. This again points at the value of simultaneously optimizing the plasma breakdown scenario with the first few seconds of plasma equilibria, since the details of plasma breakdown do affect features like time-to-divert.

\section{Conclusion and outlook}

In conclusion, the GSPulse algorithm is a novel method of solving free-boundary equilibrium evolution (FBEE) that reformulates the problem in order to efficiently perform trajectory optimization. This has a wide variety of applications for scenario development and scoping, as well as utility in providing feedforward trajectories to improve feedback control performance. GSPulse has been validated against experiments and the pulse simulator FGE. GSPulse continues to undergo active development and feature additions. Additional features that are under development include: adding the capability to optimize across the plasma breakdown transition (before plasma, and during very low-$I_p$ plasmas), adding a Python API to the code base so that GSPulse can interface with Python codes, and extending the optimization to be able to optimize for more complex error types, such as the structural forces exerted on the coils which have a nonlinear relationship with respect to coil currents. The code is available open-source at \blue{https://github.com/jwai-cfs/GSPulse\_public}. 

\newpage
\pagebreak
\vspace{1in}
\appendix

\section{GSPulse Algorithm}\label{app:gspulse_algorithm}

\subsection{Model equations:} \label{subsec:model_equations} 

This appendix describes the GSPulse algorithm in detail. To recall, the objective of GSPulse is to minimize a quadratic cost function on the actuator effort and shaping errors:

\begin{equation}
    J = \int_{\Omega_t} ||u||^2_{W_u} +  ||\dot u||^2_{W_{\dot u}} + ||\ddot u||^2_{W_{\ddot u}} + ||e||^2_{W_e} +  ||\dot e||^2_{W_{\dot e}} + ||\ddot e||^2_{W_{\ddot e}} \; dt
\end{equation}

This cost function is discretized such that the specific implementation that is optimized for is: 

\begin{equation}\label{eq:gspulse_cost}
\begin{aligned}
    J &:= \sum_{k=k_0}^{N} ||u_{k-1}||^2_{W_u} +  ||\Delta u_{k-1}||^2_{W_\Delta u} + ||\Delta^2 u_{k-1}||^2_{W_\Delta^2 u} + ||e_k||^2_{W_e} +  ||\Delta e_k||^2_{W_\Delta e} + ||\Delta^2 e_k||^2_{W_\Delta^2 e}, \\ \\
    k_0 &:= \text{1 for absolute terms, 2 for first-order difference ($\Delta$) terms,} \\ &\;\;\;\;\;\; \text{3 for second-order difference ($\Delta^2$) terms}
\end{aligned}
\end{equation}

The cost should be minimized while subject to the Grad-Shafranov equilibrium force balance condition, axisymmetric conductor circuit dynamics, and $I_p$ dynamics via the Ejima equation: 

\begin{equation}\label{A1eq:GS}
    \begin{split}
    \Delta^* \psi &= -\mu_0 R J_\phi \\
    J_\phi &:= J_{\phi}^{pla} + J_{\phi}^{vac} \\
    J_{\phi}^{pla} &= RP'(\psi) + \frac{FF'(\psi)}{\mu_0 R}
    \end{split}
\end{equation}

\begin{equation}
    u_i = R_i I_i + M_{ij} \dot I_j + \dot \Phi_i^{pla}
\end{equation}

\begin{equation}\label{eq:ejima}
    V_p = -\dot \psi_{bry} = R_p I_p + \frac{1}{I_p}  \dv{}{t} \left(\frac{1}{2}L_I I_p^2 \right)
\end{equation}

where,

\begin{equation}\label{eq:fbee_variable_defs}
\begin{split}
    \Delta t &= \text{time step resolution} \\
     u_k &= \text{actuator input (voltage) at time step k}\\
     \Delta u_k &:= (u_{k+1} - u_k) / \Delta t  = \text{actuator finite-difference first derivative}\\
    \Delta^2 u_k &:= (u_{k+2} - 2u_{k+1} + u_k) / \Delta t^2  = \text{actuator finite-difference second derivative}\\
    W_{(\cdot)} &= \text{cost function weighting matrix} \\
    e_k &= \text{output errors such as isoflux shaping and field errors at time step k} \\
    \Delta e_k &:= (e_{k+1} - e_k) / \Delta t = \text{output error first derivative}\\
    \Delta^2 e_k &:= (e_{k+2} - 2e_{k+1} + e_k) / \Delta t^2 = \text{output error second derivative}\\
     \Delta^*(\cdot) &= r\frac{\partial }{\partial r} \left( \frac{1}{r} \frac{ \partial (\cdot) }{\partial r } \right) + \frac{\partial^2 (\cdot) }{\partial z^2} = \text{Grad-Shafranov operator} \\
     \mu_0 &= \text{vacuum permeability constant} \\
     r &= \text{radial coordinate} \\
     J_\phi &= \text{toroidal current density distribution} \\
     P &= \text{pressure} \\
     \psi &= \text{poloidal flux per unit radian} \\
     F &= rB_t = \text{radius times toroidal magnetic field}. \\
     i,j &= \text{indices for conducting elements} \\
     R_i &= \text{resistance in conductor i} \\
     I_i &= \text{current in conductor i} \\
     M_{ij} &= \text{mutual inductance between conductors i and j} \\
     \dot \Phi_i^{pla} &= \text{plasma coupling term, discussed in text} \\
     V_p &= \text{plasma loop voltage} \\
     \psi_{bry} &= \text{flux at plasma boundary} \\
     R_p &= \text{total plasma resistance} \\
     I_p &= \text{total plasma current} \\
     t &= \text{time} \\
     L_I &= \int_{\Omega_p} \frac{B_p^2}{\mu_0 I_p^2}  dV \text{  plasma internal inductance (unnormalized)}
\end{split}
\end{equation}

\subsection{Algorithm Steps}

The core idea of the GSPulse algorithm is to use a Picard iteration scheme for the Grad-Shafranov plasma flux distribution coupled with a time-dependent, global, dynamic optimization of the vacuum conductors. A diagram of the convergence strategy was given in \cref{fig:gspulse_convergence}, illustrating the iterations between global optimization and plasma Picard updates of the flux distribution. After parsing the configuration inputs, the GSPulse algorithm performs the following actions:

\noindent\fbox{%
\parbox{\dimexpr\linewidth-2\fboxsep-2\fboxrule\relax}{
\begin{enumerate}[itemsep=0.5em, leftmargin=*]

\item \textbf{Initialization:} Estimate the initial plasma current distribution at each time step.  

\item \textbf{Compute target boundary flux:} Integrate the Ejima equation to find the target boundary flux.  

\item \textbf{Optimize conductor evolution:} Set up and solve a QP for the conductor evolution. 

\item \textbf{Grad-Shafranov Picard iteration:} Update the plasma flux distribution

\item \textbf{Convergence:} Repeat from Step 2 until convergence metrics are satisfied.  
\end{enumerate}
            }%
}

These steps are described in detail below:

\vspace{1cm}
\noindent \textbf{Step 1: Initialization} \newline

The purpose of this step is to obtain a rough estimate of the plasma current distribution. We use the target shape boundary and find the geometric centroid. Then each point on the grid is written in terms of a scaled distance x with x=0 at the centroid and x=1 at the boundary. The current is estimated with parabolic distribution:

\begin{equation}
    J_\phi = \hat J (1-x^a)^b
\end{equation}

where $a$ and $b$ are constants and $\hat J$ is a constant scaled to match the target $I_p$. \newline

\noindent \textbf{Step 2: Calculate target boundary flux} \newline

The plasma internal inductance $L_I$ can be measured from the current distribution. With initial condition $\psi_{bry}(t=0)$ obtained from the starting equilibrium, the Ejima \cref{eq:ejima} can be integrated to find the $\psi_{bry}(t)$ that provides the surface voltage to drive the target $I_p$. \newline

\noindent \textbf{Step 3: Optimize conductor evolution} \newline

This step updates the coil currents and applied flux, and is done by casting the update into the form of a quadratic optimization problem using a technique similar to that used in Model Predictive Control (MPC).

The first step is casting the circuit dynamics equation into state space form. The circuit equation is: 

\begin{equation}\label{A1eq:circuit}
    v_i = R_i I_i + M_{ij} \dot I_j + \dot \Phi_i^{pla} \\
\end{equation}

We also have that the plasma-induced flux at any specific conductor is related through the grid mutuals and the plasma current distribution. 

\begin{equation}\label{A1eq:phi_pla}
    \dot \Phi_i^{pla} = M_{ig} \dot I_g
\end{equation}

where $M_{ig}$ is the mutual inductance between conductor $i$ and each grid location 
$g$, and $I_g$ is the plasma current in grid cell $g$. (Note the connection to the Grad-Shafranov equation: $I_g$ is the plasma current distribution corresponding to the current density $J_\phi$).

The conductor evolution is written for both coil and vessel elements. Combining \cref{A1eq:circuit,A1eq:phi_pla} and being explicit about coils or vessel elements, we arrive at:

\begin{equation}\label{A1eq:cond_dynamics}
    \begin{bmatrix} V_c \\ 0 \end{bmatrix} = R_{cv} \begin{bmatrix} I_c \\ I_v \end{bmatrix} + M_{cv,cv} \begin{bmatrix} \dot I_c \\ \dot I_v \end{bmatrix} + M_{cv,g} \dot I_g 
\end{equation}

Re-arranging this matrix equation gives us:

\begin{equation}
    \dot x = A x + B u + w 
\end{equation}

where we have used the following substitutions

\begin{equation}    
\begin{split}   
    x &= [I_c^T \;\; I_v^T]^T \\
    u &= V_c \\
    A &= -M_{cv,cv}^{-1}R_{cv} \\
    B &= M_{cv,cv}^{-1} \begin{bmatrix} I \\ 0 \end{bmatrix} \\
    w &= -M_{cv,cv}^{-1}M_{cv,g} \dot I_g
\end{split}    
\end{equation}

Note that $w$ represents the influence of the plasma current on the conductors and can be calculated directly given the plasma current distribution at each time. Except for $w$ which depends on the Grad-Shafranov solutions and is updated each iteration, the remaining dynamic terms are constant for a given machine (i.e. $A$ and $B$ do not change across GS iterations). This dynamics model is converted to discrete-time using the zero order hold method. For conciseness, we abuse the notation by using the same labels, making note that A, B, and w from hereon refer to their discrete time versions. 

\begin{equation}\label{A1eq:discrete_dynamics}
    x_{k+1} = A x_k + B u_k + w_k
\end{equation}

The shaping targets include parameters such as the flux at each of the control points, flux at target boundary location, and field at the target x-points. For any output $y$ that is a linear function $f$ of the grid flux distribution $\psi_g$ (which is true for flux and field measurements) then the output can be represented:

\begin{equation}
    \begin{split}
        y &= f(\psi_g) \\
          &= f(\psi_g^{app}) + f(\psi_g^{pla}) \\        
          &= f(M_{g,cv}I_{cv}) + f(\psi_g^{pla}) \\
          &:= f(M_{g,cv}I_{cv}) + y^{pla}
    \end{split}
\end{equation}

In other words, the output is separated into a part that depends on the coil and vessel currents and part that depends on the plasma flux distribution. Moreover, for flux and field measurements the function $f(\cdot)$ is a linear mapping (e.g. $f(x) = Cx$ where C is the greens functions for flux or field) so that we can write

\begin{equation}
    y = CI_{cv} + y^{pla} \\
\end{equation}

or using the state-space notation

\begin{equation}\label{A1eq:output}
    y_k = C x_k + y_k^{pla}
\end{equation}

The dynamics \cref{A1eq:discrete_dynamics} and output \cref{A1eq:output} form the basis of the prediction model for the optimization. In both equations we have intentionally separated the linear and nonlinear portions. The nonlinear terms $w_k$ and $y_k^{pla}$ are measured from the plasma current distribution and updated each iteration, while other terms can be computed once and re-used across iterations. Our cost function for the optimizer was \cref{eq:gspulse_cost}, repeated here:

\begin{equation}
    J := \sum_{k=k_0}^{N} ||u_{k-1}||^2_{W_u} +  ||\Delta u_{k-1}||^2_{W_\Delta u} + ||\Delta^2 u_{k-1}||^2_{W_\Delta^2 u} + ||e_k||^2_{W_e} +  ||\Delta e_k||^2_{W_\Delta e} + ||\Delta^2 e_k||^2_{W_\Delta^2 e}    
\end{equation}

Where $u_k$ is the power supply voltage and $e_k = r_k - y_k$ is the shape error between the reference target and actual, and the $W_{(x)}$ are user-defined weights on the magnitude, derivative, and second derivative of the voltages and errors. 

The next few steps map the cost function into a standard quadratic program which can be solved by many available software packages such as MatLab's \textbf{quadprog}. The next steps aim to transform the above cost function into the standard quadratic program form,

\begin{equation}\label{A1eq:standardform}
    J = \hat p^T H \hat p + 2f^T \hat p,
\end{equation}

where $\hat p$ is the set of primary optimization variables to be solved for, and for the GSPulse problem we use: 

\begin{equation}
    \hat p := \begin{bmatrix} x_1 \\ \hat u \end{bmatrix}
\end{equation}
 
To begin, we make the following definitions:
\begin{equation}
    \begin{split}
        \hat u & := [u_0^T \; u_1^T \; ... \; u_{N-1}^T]^T  \\
        \widehat {\Delta u} & := [(u_1-u_0)^T \;\; (u_2-u_1)^T \; ... \; (u_{N-1}-u_{N-2})^T]^T / \Delta t \\
        \widehat {\Delta^2 u} & := [(u_2-2u_1+u_0)^T \;\; (u_3-2u_2+u_1)^T \; ... \; (u_{N-1}-2u_{N-2}+u_{N-3})^T]^T / \Delta t^2 \\
        \hat e & := [e_1^T \; e_2^T \; ... \; e_{N}^T]^T \\     
        \widehat{\Delta e} &:= [(e_2-e_1)^T \;\; (e_3-e_2)^T \; ... \; (e_{N}-e_{N-1})^T]^T / \Delta t\\
        \widehat{ \Delta^2 e} &:= [(e_3-2e_2+e_1)^T \;\; (e_4-2e_3+e_2)^T \; ... \; (e_{N}-2e_{N-1}+e_{N-2})^T]^T / \Delta t^2\\
        \hat w & := [ w_0^T \; w_1^T \; ... \; w_{N-1}^T]^T \\
        \hat x & := [x_1^T \; x_2^T \; ... \; x_{N}^T]^T \\        
        \hat y & := [y_1^T \; y_2^T \; ... \; y_{N}^T]^T \\        
        \hat r & := [r_1^T \; r_2^T \; ... \; r_{N}^T]^T \\                
        \hat C & := \textbf{blkdiag(} \underbrace{C, C \; ... \; C}_{\times N} \textbf{)} \\
        \hat W_{(\alpha)} &:= \textbf{blkdiag(} \underbrace{W_{(\alpha)}, W_{(\alpha)} \; ... \; W_{(\alpha)}}_{\times (N - k_0 + 1)} \textbf{)}, \;\;\;\;\;\; \alpha \in {u, \Delta u, \Delta^2 u, e, \Delta e, \Delta^2 e} \\
        k_0 & := \text{summation starting index = 1 for absolute terms, 2 for first-order difference } \\
        & \;\;\;\;\; \text{($\Delta$) terms, 3 for second-order difference ($\Delta^2$) terms}
    \end{split}
\end{equation}

Note that we can write matrix mappings between the $\hat u$ and $\Delta \hat u$, since these are just linear combinations of each other. 

\begin{equation}
    \widehat{\Delta u} = \frac{1}{\Delta t}\underbrace{\begin{bmatrix}
        -I & I \\
        & \ddots \\
        & & -I & I \\
    \end{bmatrix}}_{:=S_\Delta u} \hat u 
\end{equation}

Similarly for the 2nd derivatives, we have that:

\begin{equation}
    \widehat{\Delta^2 u} = \frac{1}{\Delta t^2} \underbrace{\begin{bmatrix}
        I & -2I & I \\
        & & \ddots \\
        & & I & -2I & I \\
    \end{bmatrix}}_{:=S_\Delta^2 u} \hat u 
\end{equation}

And for equivalently-defined transition matrices for the errors, we have that:

\begin{equation}
\begin{split}
    \widehat{\Delta e} = S_{\Delta e} \hat e \\
    \widehat{\Delta^2 e} = S_{\Delta^2 e} \hat e
\end{split}
\end{equation}

Thus the first 3 terms and last 3 terms in the cost function \cref{eq:gspulse_cost} can be combined to rewrite the cost function as a matrix equation: 

\begin{equation}\label{eq:QP2}
\begin{split}
    J &= \hat u^T H_u \hat u + \hat e^T H_e \hat e \\
    H_u &:= \hat W_u + S_{\Delta u}^T \hat W_{\Delta u} S_{\Delta u} +  S_{\Delta^2 u}^T \hat W_{\Delta^2 u} S_{\Delta^2 u} \\
    H_e &:= \hat W_e + S_{\Delta e}^T \hat W_{\Delta e} S_{\Delta e} +  S_{\Delta^2 e}^T \hat W_{\Delta^2 e} S_{\Delta^2 e} \\
\end{split}
\end{equation}

Now, lets turn our attention to the dynamics to ``unroll'' or project forward how future states depend on the sequence of actuator actions. This step is adapted from MPC, where the unrolled system is called the prediction model in MPC contexts. From the state-space dynamics equation we have that: 

\begin{equation}
\begin{split}
    x_1 &= x_1 \\
    x_2 &= A x_1 + B u_1 + w_1 \\
    x_3 &= A x_2 + B u_2 + w_2 \\
        &= A(x_1 + B u_1 + w_1) + B u_2 + w_2 \\
        &= A^2 x_1 + \begin{bmatrix} AB & B \end{bmatrix} \begin{bmatrix} u_1 \\ u_2 \end{bmatrix} + \begin{bmatrix} A & I \end{bmatrix} \begin{bmatrix} w_1 \\ w_2 \end{bmatrix}\\
    x_4 &= ...\\
\end{split}
\end{equation}

Extending this to all times gives: 

\begin{equation}  
    \begin{bmatrix} x_{1} \\ x_{2} \\ \vdots \\ x_{N} \end{bmatrix} 
    =
    \underbrace{\begin{bmatrix} I \\ A \\ \vdots \\ A^N \end{bmatrix}}_{E} x_1
    +
    \underbrace{\begin{bmatrix} 0 \\ 0 & B \\ 0 & AB & B \\ \vdots & \vdots & \ddots \\  0 & A^{N-1}B &  A^{N-2}B & \hdots & B \end{bmatrix}}_{F}
    \begin{bmatrix} u_0 \\ u_{1} \\ \vdots \\ u_{N-1} \end{bmatrix}
    + \underbrace{\begin{bmatrix} 0  \\  0 & I \\  0 & A & I \\ \vdots & \vdots & \ddots \\  0 & A^{N-1} &  A^{N-2} & \hdots & I \end{bmatrix}}_{F_w}
    \begin{bmatrix}  w_0 \\ w_1 \\ \vdots \\ w_{N-1} \end{bmatrix}    
\end{equation}

Or equivalently,

\begin{equation}\label{eq:pm}
    \hat x = Ex_1 + F \hat u + F_w \hat w
\end{equation}

Then using \cref{A1eq:output} the predicted errors are:

\begin{equation}\label{A1eq:epm}
\begin{split}
    \hat e &= \hat r - \hat y \\
        &= \hat r - \hat y^{pla} - \hat C \hat x \\
        &= \hat r - \hat y^{pla} - \hat C (Ex_1 + F \hat u + F_w \hat w) \\
        &:=  M \hat p  + d 
\end{split}    
\end{equation}

where to obtain the last line we have defined new variables,

\begin{equation}\label{eq:mpd}
\begin{split}
    M &:= -\hat C \begin{bmatrix} E & F \end{bmatrix} \\
    \hat p &:= \begin{bmatrix} x_1 \\ \hat u \end{bmatrix} \\
    d &:= \hat r - \hat y^{pla} - \hat C F_w \hat w
\end{split}
\end{equation}

Note that $\hat p$ consists of the initial starting currents $x_1$ and the voltages at all times, and is the set primary optimization variables to be solved for by the quadratic program \cref{A1eq:standardform}. That is, the output of the QP solve is $\hat p$. The elements of $x_1$ can be specified either as constrained parameters, for the use case where an initial condition is specified, or they can be specified as free parameters for the optimizer to solve for. 

The relation between the voltages and the primary optimization variables $\hat p$ is: 

\begin{equation}\label{eq:Tup}
    \hat u = \underbrace{\begin{bmatrix} 0 & \\ & I \end{bmatrix}}_{:= T_{up}} \hat p
\end{equation}

Then, plugging in \cref{eq:Tup,A1eq:epm} into \cref{eq:QP2}, we arrive finally at the standard form of the quadratic program: 

\begin{equation}\label{eq:QP}
\begin{split}
    J &= \hat u^T H_u \hat u + \hat e^T H_e \hat e \\
    &= \hat p^T T_{up}^T H_u T_{up} \hat p + (M \hat p  + d)^T H_e (M \hat p  + d) \\
    &= \hat p^T H \hat p + 2f^T \hat p \\
    H &:= T_{up}^T H_u T_{up} + M^T H_e M \\
    f &:= M^T H_e^T d
\end{split}
\end{equation}

Solving the quadratic program \cref{eq:QP} can be done with standard QP solver packages such as \textbf{quadprog} in MATLAB. The solution to the QP is the initial starting currents as well as the trajectory of voltages. The voltage trajectory can be mapped back to the trajectory of currents and errors, via \cref{eq:pm} and \cref{A1eq:epm}. Constraints on the voltages, currents, and outputs can be added to the QP problem via the mappings presented in this derivation. In the standard QP form, constraints enter in the form of $A \hat p < b$, which is a linear mapping of the primary optimization variables. The following equations describe how to put various types of constraints in this form. 

For constraints on the voltage ($\hat u$), which would include restrictions such as power supply voltage limits or slewrate limits,  we use \cref{eq:Tup} to write:

\begin{equation}\label{eq:QP_ineq1}
\begin{split}
    & A_u \hat u < b_u \\
    \implies & A_u T_{up} \hat p < b_u
\end{split}
\end{equation}

Similarly for constraints on the coil currents ($\hat x$) we use \cref{eq:pm} to write: 
\begin{equation}\label{eq:QP_ineq2}
\begin{split}
    &A_x \hat x < b_x \\ 
    \implies &A_x (E x_1 + F \hat u + F_w \hat w) < b_x \\
    \implies &A_x ([E\; F] \hat p + F_w \hat w) < b_x \\
    \implies &A_x [E \; F] \hat p < b_x - A_x F_w \hat w
\end{split}
\end{equation}

And for constraints on the output errors ($\hat e$) we use \cref{A1eq:epm} to write: 

\begin{equation}\label{eq:QP_ineq3}
\begin{split}
    & A_e \hat e < b_e \\
    \implies & A_e (M \hat p + d) < b_e \\
    \implies & A_e M \hat p < b_e - A_e d
\end{split}
\end{equation}

Lastly, we conclude with a few observations about this QP. 

\begin{itemize}
    \item Note that, if only 1 time step is being solved for ($N=1$), then the dynamical prediction model shrinks to identity and the QP only solves for the $x_1$ set of currents. In other words, it behaves exactly as a static inverse Picard-based equilibrium solver updating a single set of currents at each iteration. This means that, in addition to being a dynamic equilibrium trajectory solver, GSPulse can also function as a standard inverse equilibrium solver, since the static problem is a sub-problem of the dynamic one. 

    \item Assuming that the state-space output matrix $\hat C$ is linear, then most terms within the QP problem formulation (\cref{eq:QP,eq:QP_ineq1,eq:QP_ineq2,eq:QP_ineq3}) are constant across iterations. The only nonlinear term that must be updated between iterations is $d = \hat r - \hat y^{pla} - \hat C F_w \hat w$ (\cref{eq:mpd}). This means that most terms within the QP can be computed only once and re-used, saving on computational time since a large fraction of compute is absorbed by obtaining the linearization and multiplying large matrices to form the QP. Therefore it is very advantageous to ensure that the output matrix $\hat C$ is linear and does not need to be updated. This is the motivation within GSPulse for specifying the error signals in terms of isoflux-type outputs (flux and field differences) instead of spatial gap-related units. For outputs that have units of B-field, flux, or current, then $\hat C$ is linear because of the linearity of the Green's functions.
\end{itemize}

\vspace{1em} \noindent \textbf{Step 4: Grad-Shafranov Picard iteration} \newline

In this step, the goal is to perform a Picard update the total flux distribution according to the Grad-Shafranov equation in order to converge the solution numerically. The full Picard iteration process is given as follows.

In the QP optimizer step, we computed the trajectory currents in the external conductors. (Actually, the QP solver obtained the initial currents and trajectory of voltages, but these are mapped to the conductor currents via the prediction model \cref{eq:pm}.) This allows us to update the vacuum flux distribution via:

\begin{equation}
    \psi_{vac}^{k+1} = M_{g,cv} I_{cv}^{k+1},
\end{equation}

where $k$ is the iteration index. Next, we do a partial update of the total flux distribution, taking the partially-updated flux as a sum of the previous plasma flux contribution and the updated vacuum flux. 

\begin{equation}\label{eq:psi_khalf}
    \psi^{k+1/2} := \psi_{pla}^k + \psi_{vac}^{k+1}
\end{equation}

The remaining steps are to update the $P'$ and $FF'$ profiles and total flux of the Grad-Shafranov equation: 

\begin{equation}\label{eq:GS_picard}
\begin{split}
    J_{\phi,pla}^{k+1} &= RP'(\psi^{k+1/2}) + \frac{FF'(\psi^{k+1/2})}{\mu_0 R} \\
    \Delta^* \psi^{k+1} &= -\mu_0 R \left( J_{\phi, pla}^{k+1} + J_{\phi,vac}^{k+1} \right)
\end{split}
\end{equation}

The present implementation of GSPulse allows for two different methods of constraining the $P'$ and $FF'$ profiles used in the Picard update. The first method is that GSPulse has been integrated with the FBT code developed by EPFL that is part of the open-source MEQ toolbox. FBT contains a number of flexible and fast methods for constraining the $P'$ and $FF'$ profiles, including various basis function types for the profile shapes, and multiple options to specify combinations of core plasma properties such as the total plasma current $I_p$, total stored thermal energy $W_{th}$, plasma beta $\beta_p$, or q-profile value on-axis $q_A$. 

The second method is a built-in GSPulse option that does not require installation of the FBT/MEQ codes, but has slightly less flexibility in specification of the core plasma properties. With this method, the user must specify input basis function shapes for the $P'$ and $FF'$ profiles, as well as input waveforms for the plasma current $I_p$ and thermal energy $W_{th}$. At each iteration, GSPulse scales the coefficients multiplying the basis function profiles in order to match the target $I_p$ and $W_{th}$. 

To describe the built-in method with more detail: 

After using the optimizer to obtain the updated conductor currents, we compute $\psi_{k+1/2}$ from \cref{eq:psi_khalf} which is the partially-updated flux on the grid. Then we use touch-point finder, x-point finder, and o-point finder algorithms on this flux distribution. Out of the identified potential touch-points, x-points, and o-points, we select the magnetic axis and boundary-defining point of the flux distribution (where the boundary-defining point is either a touch-point or an x-point). The algorithm to identify which points are the magnetic axis and boundary follows the logic presented in \cite{Moret2015} section 2.3, which is an efficient method for identifying the boundary. The output of all these steps is to identify the flux at the magnetic axis $\psi_{mag}$ and the flux at the boundary $\psi_{bry}$ as well as to identify the physical boundary of the plasma. 

Since the user has provided input profile shapes on a normalized $\psi$ basis, our job is to appropriately scale the coefficients in order to match the target $I_p$ and $W_{th}$. Let the input basis shapes be $P'_b({\psi_N})$ and $FF'_b({\psi_N})$ where $\psi_N$ is the normalized flux:

\begin{equation}
    \psi_N := \frac{\psi - \psi_{bry}}{\psi_{mag} - \psi_{bry}}
\end{equation}

The total profiles are then: 

\begin{equation}
\begin{split}
    P'(\psi) &= c_p P'_b(\psi_N) \; / \; (\psi_{mag} - \psi_{bry}) \\
    FF'(\psi) &= c_f FF'_b(\psi_N) \; / \; (\psi_{mag} - \psi_{bry}) \\
\end{split}
\end{equation}

The scaling coefficients $c_p$ and $c_f$ are terms to be solved for, and the last term arises from de-normalizing the $\psi$ basis, since for any parameter $x$:

\begin{equation}
\begin{split}
    \frac{dx}{d\psi} &= \frac{dx}{d\psi_N} \frac{d \psi}{d \psi_N} \\
    &= \frac{dx}{d\psi_N} \frac{1}{(\psi_{mag} - \psi_{bry})}
\end{split}
\end{equation}

The first constraint is on the total plasma current,

\begin{equation}
    I_p = \int_{\Omega_p} J_{\phi, pla} dA
\end{equation}

Substituting in the Grad-Shafranov \cref{A1eq:GS}, we obtain that

\begin{equation}\label{A1eq:Ip}
    I_p = \frac{1}{\psi_{mag}-\psi_{bry}} \left [ c_{p} \int_{\Omega_p} RP'_b dA  + c_{f} \int_{\Omega_p} \frac{FF'_{b}}{\mu_0 R} dA \right ]
\end{equation}

This equation depends linearly on the profile coefficients, while the integrals are computed numerically at each iteration. The second constraint is on the stored thermal energy $W_{th}$, where we have that:

\begin{equation}\label{A1eq:Wth}
\begin{split}
    W_{th} &= \int_{\Omega_p} \frac{3}{2} P(\psi_N) dV \\   
    &= \int_{\Omega_p} 3 \pi R P(\psi_N) dA
\end{split},
\end{equation}

where the pressure is found from integrating the $P'$ basis function from the edge of the plasma inwards (assuming that the pressure is zero at the edge). 

\begin{equation}\label{eq:p_constraint}
    P(\psi_N) = c_{p} \int_{\hat \psi_N=1}^{\hat \psi_N = \psi_N} P'_b d \hat \psi_N
\end{equation}

The linear system of 2 equations (\cref{eq:p_constraint,A1eq:Ip}) and 2 unknowns can be solved to obtain the profile coefficients. This gives us the necessary total $P'$ and $FF'$ profiles, which are used to obtain the total flux distribution according to \cref{eq:GS_picard}. 

At this stage we have now updated the plasma current and total flux distributions, and the solver can proceed to the next iteration.

\subsection{Spline basis compression:} 

The computational complexity of obtaining an optimal solution to the QP (\cref{eq:QP}) can vary dramatically depending on the number of equilibria that are solved for and the complexity of the evolution. Solving this type of optimization problem can have poor numerical scaling, since in general the number of optimization variables that is solved for is $n = (N_{equlibria}+2) * N_{coils}$, and the theoretical time complexity for solving linear-constrained quadratic programs is $O(n^3)$. In practice the run-time performance also depends on the complexity of the equilibrium evolution. Observationally, highly dynamic events such as strike point sweeping take more time to solve. Therefore it is advantageous to employ methods that reduce the computational complexity of the QP perhaps at the expense of sacrificing some of the optimality of the solution. 

One technique for data compression is to use a spline basis for the primary optimization variables. If we have a spline basis of the form,

\begin{equation}\label{eq:spline}
    \hat p = S_m \hat c
\end{equation}

where $\hat p$ is the original primary variables, $S_m$ is a splining matrix, and $\hat c$ are the spline basis coefficients. Then, instead of solving for the original variables we optimize for the spline coefficients. The quadratic program is then transformed as:

\begin{equation}
\begin{split}
    \text{min } J &= \hat p^T H \hat c + 2f^T \hat p \\
                &= \hat c^T S_m^T H S_m \hat c + 2 f^t S_m \hat c \\
    \text{subj to: } \\
    & A \hat p \leq b \\
    \implies & A S_m \hat c \leq b
\end{split}    
\end{equation}

After solving for the spline coefficients, the original variables are constructed from the spline basis above \cref{eq:spline}. 

\subsection{Optimization with multiple time intervals:}

Another important technique for reducing the time complexity of the QP-solve is to solve for the evolution in multiple stages. For example, instead of solving for 200 equilibria at a time, we may solve for 50 equilibria at a time across 4 overlapping stages reducing the complexity from $O(200^3)$ to $4\times O(50^3)$. This reduces the optimality of the solve since the first stage is not looking far into the future. However, in many cases this optimality reduction is not significant, i.e. details of the rampdown trajectory do not significantly affect the details of the rampup trajectory. 

GSPulse enables this multiple-interval approach by overlapping the equilibria from one stage to the next. The $1^{st}$ and $2^{nd}$ equilibria are automatically enforced to be the $(N-1)^{th}$ and $N^{th}$ equilibria from the previous stage. These equality constraints enforce continuity of the trajectory, and also give smooth trajectories provided that the user has weighting penalties on the second derivative (non-smooth trajectories). 

\begin{figure}[H]
    \begin{center}
        \includegraphics[width=12cm]{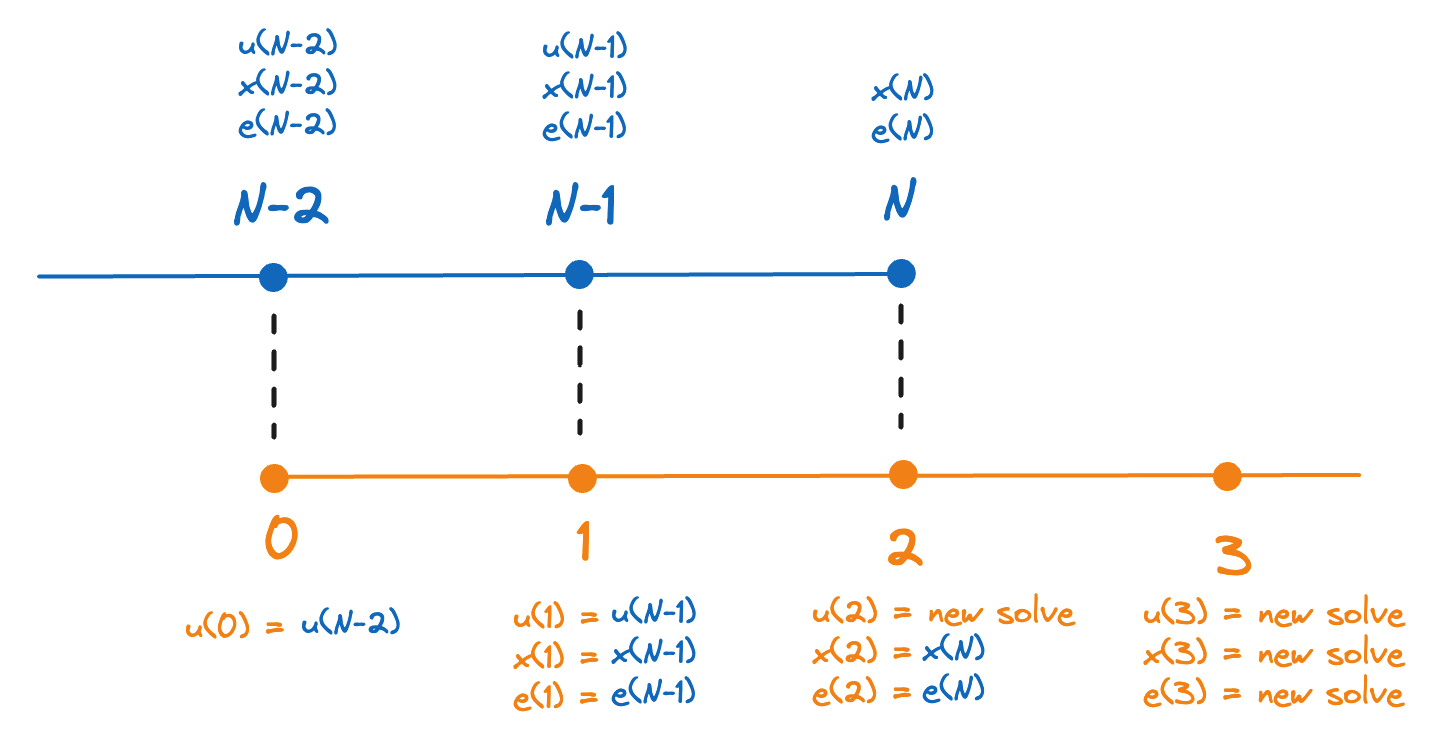}
    \end{center}
    \caption{Overlapping strategy when using GSPulse to optimize across multiple intervals. The previous stage is shown in blue and subsequent stage in orange. The stages are overlapped by 2 equilibria (2 time points each for the voltages $u$, currents $x$, and shaping errors $e$).}
    \label{fig:blend_stages}
\end{figure}

The multiple-interval strategy is shown in \cref{fig:blend_stages}. As shown in the figure, the subsequent stage needs to enforce equality constraints on $u_0, u_1, x_1, e_1, x_2$ and $e_2$. 

\begin{itemize}
    \item The terms $x_1, u_0$ and $u_1$ make up the first three terms of the optimization variables $\hat p$ (see \cref{eq:mpd}). Therefore, these terms can be enforced by equality constraint on the QP solver. 

    \item The equality constraint on $x_2$ is enforced automatically and implicitly through the other terms, since $x_2 = Ax_1 + Bu_1 + w_1$ and all terms on the RHS are specified. 

    \item The remaining terms are $e_1$ and $e_2$. Our prediction model for the errors was \cref{A1eq:epm} and repeated here: 

    \begin{equation}\label{eq:epm}
        \hat e := \begin{bmatrix} e_1 \\ e_2 \\ e_3 \\ \vdots \\ e_N \end{bmatrix} = M \hat p + d
    \end{equation}

    We achieve equality constraint by replacing the rows of $M$ and elements of $d$ corresponding to $e_1$ and $e_2$:

\begin{equation}\label{eq:epm_stage}
\begin{bmatrix} e_1 \\ e_2 \\ e_3 \\ \vdots \\ e_N \end{bmatrix} = 
\underbrace{\left [\begin{array}{c}
	0 \\ 0\\ 
	\hline 
	 \\ M_{((2\times n_e + 1):N,:)} \\ \\
\end{array} \right ]}_{\bar M} \hat p + 
\underbrace{\left [\begin{array}{c}
	e_1 \\ e_2\\ 
	\hline 
	 \\ d_{(2\times n_e + 1):N} \\ \\
\end{array} \right ]}_{\bar d}
\end{equation}

    Then, \cref{eq:epm_stage} is used in place of \cref{eq:epm} when constructing the QP, for any stage after the first.
\end{itemize}

\pagebreak


\section*{Acknowledgements}
This material is based upon work supported by the U.S. Department of Energy, Office of Science, Office of Fusion Energy Sciences, under Awards DE-AC02-09CH11466, DE-SC0015480, DE-SC0021275, and DE-SC0015878.

This work has been part-funded by the EPSRC Energy Programme [grant number EP/W006839/1]. To obtain further information on the data and models underlying this paper (in Section 3.2) please contact PublicationsManager@ukaea.uk. 

This work was supported by Commonwealth Fusion Systems. 

\section*{Disclaimer}
This report is prepared as an account of work sponsored by an agency of the United States Government. Neither the United States Government nor any agency thereof, nor any of their employees, makes any warranty, express or implied, or assumes any legal liability or responsibility for the accuracy, completeness, or usefulness of any information, apparatus, product, or process disclosed, or represents that its use would not infringe privately owned rights. Reference herein to any specific commercial product, process, or service by trade name, trademark, manufacturer, or otherwise, does not necessarily constitute or imply its endorsement, recommendation, or favoring by the United States Government or any agency thereof. The views and opinions of authors expressed herein do not necessarily state or reflect those of the United States Government or any agency thereof.

\bibliography{bib}

\end{document}